\documentclass[9pt,letterpaper]{article}
\usepackage[letterpaper, total={7.2in, 9.8in}]{geometry}
\usepackage{graphicx}
\usepackage{helvet}
\usepackage{authblk}
\usepackage{hyperref}
\usepackage{amsmath} 
\usepackage{amssymb} 
\usepackage{orcidlink} 
\usepackage{booktabs} 
\usepackage[numbers,super,sort,compress]{natbib}
\usepackage{hyperref}
\usepackage{url}
\hypersetup{
	colorlinks=true,
	linkcolor=blue,
	filecolor=magenta,
	urlcolor=blue,
	citecolor=blue,
	pdftitle={HMLMs},
	pdfauthor={Hasi Hays},
}
\makeatletter
\providecommand\doi[1]{\href{https://doi.org/#1}{\nolinkurl{#1}}}
\makeatother
\makeatletter
\renewcommand{\maketitle}{\bgroup\setlength{\parindent}{0pt}
	\begin{flushleft}
		\textbf{\@title}
		
		\@author
	\end{flushleft}\egroup}
\makeatother

\usepackage{caption}
\newcounter{suppfigure}
\usepackage{changepage}
\usepackage{tikz}
\usepackage{amsmath}
\usepackage{amssymb}
\usepackage{xcolor}
\usepackage[normalem]{ulem}

\usetikzlibrary{shapes,arrows,positioning,fit,backgrounds}

\usepackage{graphicx}%
\usepackage{multirow}%
\usepackage{amsmath,amssymb,amsfonts}%
\usepackage{amsthm}%
\usepackage{mathrsfs}%
\usepackage[title]{appendix}%
\usepackage{xcolor}%
\usepackage{textcomp}%
\usepackage{manyfoot}%
\usepackage{booktabs}%
\usepackage{algorithm}%
\usepackage{algorithmicx}%
\usepackage{algpseudocode}%
\usepackage{listings}%
\usepackage{tikz}
\usetikzlibrary{shapes,arrows,positioning,fit,calc,matrix,decorations.pathmorphing,decorations.markings}
\usepackage{pgfplots}
\usepackage{booktabs}
\usepackage{changepage}
\usepackage{multicol}
\usepackage{balance}
\usepackage{float}
\usepackage{titlesec}
\usepackage{caption}
\usepackage{hyperref}
\usepackage[noabbrev]{cleveref}

\titleformat{\section}
{\normalsize\bfseries}
{\normalsize\thesection}
{1em}
{}

\titleformat{\subsection}
{\normalsize\bfseries}
{\normalsize\thesubsection}
{1em}
{}
\let\oldsubsubsection\subsubsection
\renewcommand{\subsubsection}[1]{\oldsubsubsection{#1}}

\usepackage{titlesec}
\titlespacing*{\section}{0pt}{12pt}{6pt}

\title{
	\begin{center}
		\rule{\linewidth}{2pt}\\[0.5cm]
		{\Large \textbf{Hierarchical Molecular Language Models (HMLMs)}}\\[0.3cm]
		\rule{\linewidth}{1pt}\\
		\vspace{0.7cm}
	\end{center}
}
\date{}

\author[1,*\orcidlink{0000-0003-0843-050X}] {\textbf{Hasi Hays}}
\author[2\orcidlink{0000-0002-9150-3986}] {\textbf{Yue Yu}}
\author[1\orcidlink{0000-0001-8678-9716}] {\textbf{William J. Richardson}}
\affil[1]{\footnotesize Department of Chemical Engineering, University of Arkansas, Fayetteville, AR 72701, USA}
\affil[2]{\footnotesize Department of Mathematics, Lehigh University, Bethlehem, PA 18015, USA}

\begin{document}
	
	\maketitle
	
\begin{adjustwidth}{0.5in}{0.5in}
\section*{\normalsize Abstract} 
Artificial intelligence (AI) is reshaping computational and network biology by enabling new approaches to decode cellular communication networks. We introduce Hierarchical Molecular Language Models (HMLMs), a novel framework that models cellular signaling as a specialized molecular language, where signaling molecules function as tokens, protein interactions define syntax, and functional consequences constitute semantics. HMLMs employ a transformer-based architecture adapted to accommodate graph-structured signaling networks through information transducers, mathematical entities that capture how molecules receive, process, and transmit signals. The architecture integrates multi-modal data sources across molecular, pathway, and cellular scales through hierarchical attention mechanisms and scale-bridging operators that enable information flow across biological hierarchies. Applied to a complex network of cardiac fibroblast signaling, HMLMs outperformed traditional approaches in temporal dynamics prediction, particularly under sparse sampling conditions. Attention-based analysis revealed biologically meaningful crosstalk patterns, including previously uncharacterized interactions between signaling pathways. By bridging molecular mechanisms with cellular phenotypes through AI-driven molecular language representation, HMLMs establish a foundation for biology-oriented large language models (LLMs) that could be pre-trained on comprehensive pathway datasets and applied across diverse signaling systems and tissues, advancing precision medicine and therapeutic discovery.

\section*{\normalsize Keywords}
Artificial intelligence, LLMs, Network modeling, Molecular language, Precision medicine
\end{adjustwidth}

\let\thefootnote\relax
\footnotetext{\hspace*{-7.4mm}%
	\footnotesize
	\texttt{$^{*}$Correspondence: \textcolor{blue}{hasih@uark.edu}}
}

\begin{multicols}{2}[\section{INTRODUCTION}\label{sec1}]
	
Large language models (LLMs) have demonstrated remarkable capabilities in processing complex sequential and relational data across diverse domains, suggesting that language-based architectures fundamentally capture principles of information processing applicable beyond natural language. We hypothesize that cellular signaling networks, which encode biological information through molecular interactions and temporal dynamics, can be effectively modeled as a specialized form of molecular language amenable to transformer-based learning frameworks. Cellular signaling networks are complex systems that process and relay information essential for cellular responses to environmental cues. Traditional modeling approaches, including ordinary differential equations (ODEs), agent-based models, Boolean networks, and statistical methods, have been integral in understanding these networks but often fail to adequately capture the nuances of context-dependent signaling, crosstalk between pathways, and the temporal dynamics inherent in biological processes. For instance, while ODEs enable detailed mechanistic modeling, they require extensive parameterization that can limit their applicability to larger networks \cite{Womack2024,Holtzapple2024}. Conversely, Boolean networks simplify molecular interactions to binary states, which compromises their granularity \cite{Ullanat2025,Voit2023}. Bayesian networks, while useful for identifying probabilistic relationships, can struggle to incorporate feedback mechanisms common in signaling pathways \cite{Voit2023}. Recent literature emphasizes the essential temporal and spatial dynamics of signaling interactions, necessitating a reassessment of traditional modeling frameworks \cite{Fages2024,Xu2024}. Recent developments in network biology emphasize a hierarchical organization that can better encapsulate the complexity of biological systems, encouraging integration of hierarchical information into network analyses \cite{Chang2024,Pinto2023}. This alignment of hierarchical structures with network representation can unveil multi-scale patterns and causal relationships. Moreover, engineering LLMs, especially their application to biological data, has paved the way for novel frameworks. These LLMs, which rely on transformer architectures, demonstrate proficiency in processing complex datasets and generating meaningful predictions, which can be adapted to understand the signaling behaviors in cellular networks \cite{Kolpakov2022,Levine2023}. Current explorations have indicated that biological sequences, analogous to linguistic structures, can leverage LLM methodologies to unravel relationships and emergent biological properties, indicating a promising cross-disciplinary synergy \cite{Huang2024,Giannantoni2024}. In light of this, we propose a novel framework, hierarchical molecular language models (HMLMs), designed to encapsulate the complexity of cellular signaling networks while utilizing advanced LLM capabilities. HMLMs aim to overcome the limitations of previous modeling approaches through multiple innovations. Specifically, HMLMs integrate multi-modal data sources and employ a graph-structured attention mechanism accommodating the intricate topology of signaling networks \cite{Tripathi2024}. By adding temporal embeddings, HMLMs can better match the actual timing of signaling events, which greatly improves the accuracy of their models \cite{Pinto2023}. The hierarchical representation within HMLMs permits them to learn at various organizational levels, from individual molecular interactions to larger pathway modules, thus enabling the integration of diverse experimental data and comprehensive analysis of signaling dynamics \cite{chen2025,Topsakal2023}. 

As cellular signaling networks are inherently complex, a systems biology approach that integrates both computational and experimental data is crucial for unraveling the intricacies of cell behavior. Combining data from various sources such as genomics, proteomics, nutrigenomic, and metabolomics enhances the ability to build a comprehensive understanding of signaling pathways \cite{Wu2025,Yetgin2025,Hays2023}. This data integration allows researchers to explore how different molecular players interact within the context of the entire cellular system, ultimately aiding in the identification of novel therapeutic targets. Leveraging these sophisticated techniques  improves the accuracy of predictive models and facilitates the study of dynamic processes that underpin cellular decisions, making it possible to simulate how cells respond to internal and external stimuli over time. Moreover, the exploration of hierarchical representations within signaling networks can illuminate the multi-scale organization of biological systems and their functional implications. Identifying substructures and modules within a network allows for the modeling of emergent behaviors that are essential for understanding complex cellular phenomena. The application of machine learning approaches to these hierarchical models further streamline the analysis of high-dimensional data, yielding new insights into the regulation of signaling pathways and their dysregulation in disease states. By prioritizing a systems-level perspective and fostering collaboration across disciplines, researchers can significantly advance efforts in precision medicine, thus enabling tailored therapeutic strategies that leverage the unique cellular contexts of individual patients. Furthermore, the adaptable nature of computational techniques from language modeling suggests that the principles underlying these approaches can effectively translate into biological contexts. Several studies have successfully applied LLM techniques to various aspects of molecular biology, revealing critical insights that traditional methods often overlook \cite{Levine2023,Yu2023}. The structured representation of data as ``molecular language”, characterized by unique tokens for signaling molecules and defined syntax for molecular interactions, provides a framework where complex relationships within signaling networks can be mapped and analyzed \cite{Si2024}. As the realms of computational systems biology and artificial intelligence increasingly converge, HMLMs represent a significant advancement in the modeling of cellular signaling networks. The capacity of these models to predict cellular responses to perturbations such as drug treatments or genetic modifications positions them as valuable tools in therapeutic discovery and personalized medicine \cite{Wu2022,Yu2023}.

This study makes several key contributions to the field of systems biology and its advancement in artificial intelligence (AI), including (1) We introduce cellular signaling as a form of molecular language with its unique grammar and semantics. As such, a theoretical foundation of molecular artificial intelligence (MAI) is established, which applies language modeling techniques to signaling networks. (2) We introduce HMLMs, a new computational architecture that adapts the transformer architecture to model signaling networks as information-processing systems across multiple scales. (3) We introduuce molecular-based information transducer mechanism to effectively capture context-dependent signaling behavior of signaling networks. (4) We show how HMLMs can be used to gain mechanistic insights into signaling dynamics, identify critical network nodes, and predict the effects of perturbations, facilitating the development of targeted therapeutic interventions. (5) We propose HMLMs as a potential model biology-oriented LLMs. In conclusion, while traditional signaling network modeling approaches provide a foundational understanding of cellular processes, the integration of hierarchical representations and advancements in language modeling offer a new foundation to enhance predictive capabilities and contextual understanding within these complex biological systems and develop biology-based AI, molecular-based MAI, or integrated AI models.

\captionsetup[figure]{labelformat=default}
\begin{figure*}[!ht]
	\includegraphics[width=1\textwidth]{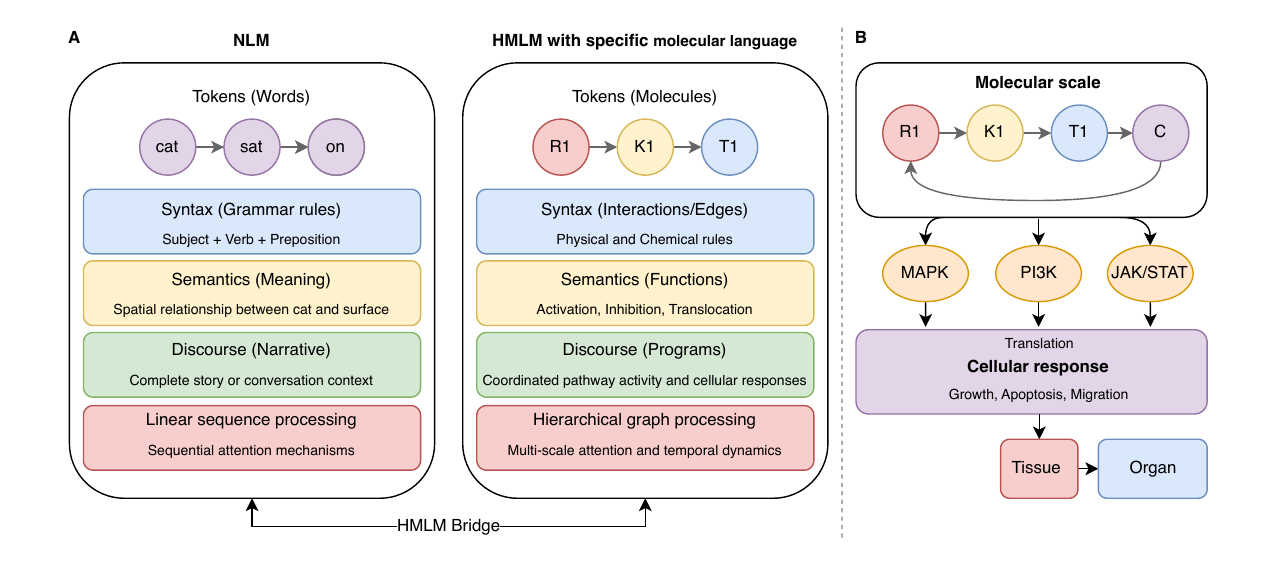}
	\caption{\footnotesize \textbf{HMLM structure and its adaptive scale representation.}}
	\footnotesize \textbf{(A) NLM and HMLM structure:} Traditional language models process linear sequences of tokens (words) through syntax (grammar rules), semantics (meaning), and discourse (narrative context) using sequential attention mechanisms. The HMLM bridge represents the principled mathematical formulation that enables this cross-domain adaptation, incorporating graph-structured attention, temporal dynamics, and scale-bridging operators for comprehensive cellular signaling network modeling. The HMLM framework maps biological signaling components to language model elements, where individual molecules (R1: receptors, K1: kinases, T1: transcription factors, C: protein complex) serve as tokens, molecular interactions define syntax through physical and chemical rules, functional consequences (activation, inhibition, translocation) constitute semantics, and coordinated pathway activities represent discourse. Unlike traditional language models, HMLMs accommodate graph structures and multi-scale organization through hierarchical architecture with specialized attention mechanisms. \textbf{(B) HMLM hierarchical scale adaptation:} The framework operates across multiple biological scales: molecular scale (individual signaling molecules), pathway scale (MAPK and other signaling pathways), and cellular scale (integrated responses like growth, apoptosis, and migration), with potential extension to tissue and organ levels. 
	\label{Fig. 1}
\end{figure*}

\begin{figure*}[!hb]
	\includegraphics[width=1\textwidth]{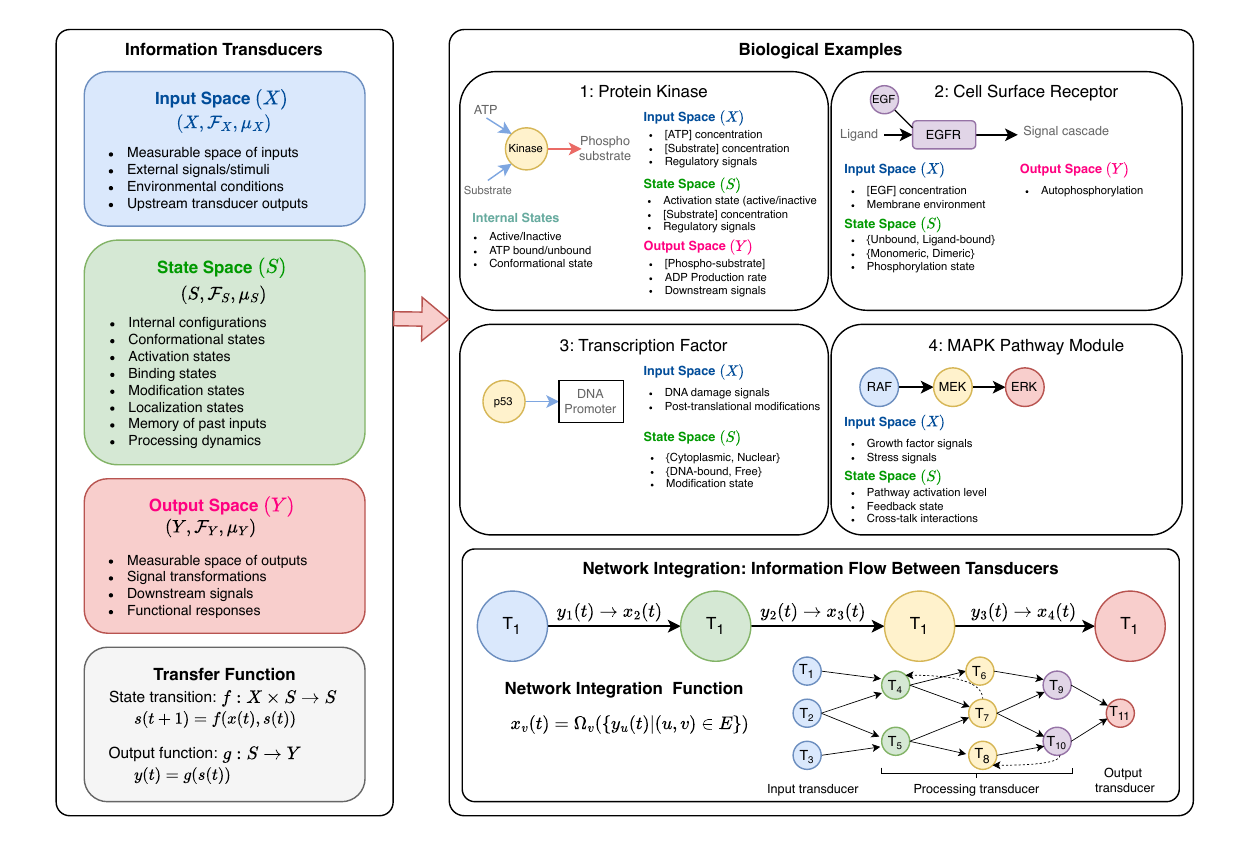}
	\caption{\footnotesize \textbf{Information transducer: Mathematical spaces to biological reality.}}
	\footnotesize The HNLM framework conceptualizes biological entities as information transducers operating within three mathematical spaces. \textbf{Mathematical Framework (Left):} Each transducer $T = (X, Y, S, f, g)$ is formally defined by input space $X$ (measurable space of external signals and stimuli), state space $S$ (internal configurations encompassing conformational, activation, binding, modification, and localization states), and output space $Y$ (measurable space of signal transformations and functional responses). State transitions follow $s(t+1) = f(x(t), s(t))$ while outputs are determined by $y(t) = g(s(t))$, with stochastic extensions using conditional probability distributions $p(s(t+1)|x(t), s(t))$ and $p(y(t)|s(t))$. \textbf{Biological Examples:} Four representative biological entities demonstrate space mapping: (1) \textbf{Protein Kinase} - input space includes ATP and substrate concentrations, state space captures active/inactive conformations and binding states, output space represents phosphorylated substrates; (2) \textbf{Cell Surface Receptor} - input space encompasses ligand concentrations, state space includes bound/unbound and monomeric/dimeric configurations, output space comprises autophosphorylation and downstream signaling; (3) \textbf{Transcription Factor} - input space contains DNA damage signals and post-translational modifications, state space captures cytoplasmic/nuclear localization and DNA binding states, output space represents transcriptional activation; (4) \textbf{MAPK Pathway Module} - input space includes growth factor and stress signals, state space encompasses pathway activation levels and feedback states, output space represents coordinated cascade responses. \textbf{Network Integration (Bottom):} Individual transducers connect through information flow equations $x_v(t) = \Omega_v(\{y_u(t)|(u,v) \in E\})$, where integration function $\Omega_v$ combines upstream outputs into downstream inputs, enabling complex network-level signal processing and emergent behaviors across multiple biological scales.
	\label{Fig. 2}
\end{figure*}

\section{METHODOLOGY}\label{sec2}

\subsection{Theoretical framework of HMLMs}

The HMLM framework bridges the gap between language modeling and biological signaling networks through a principled mathematical formulation. It is one of the paths to develop AI in medicine. This approach creates cellular signaling as a specialized ``molecular language'' where information is encoded, transmitted, and transformed across multiple scales of biological organization (\autoref{Fig. 1}). At its core, the HMLM framework remodels the signaling networks as a complex information processing system with hierarchical structure. Individual signaling molecules (proteins, metabolites, and ions) as tokens (nodes, or the basic units of information), their interactions act as syntax (edges, or the specific physical and chemical rules), and the functional consequences of these interactions constitute the meaning as semantics (activation, inhibition, translocation), and the coordinated activity of multiple pathways represents complete cellular programs as discourse (representing the narratives). This model enables us to adapt the transformer architecture of large language models to the specific challenges of modeling signaling networks. Unlike traditional language models that process linear sequences of tokens, HMLMs must accommodate the graph structure of signaling networks, the multi-scale nature of biological organization, and the temporal dynamics of signaling events. We address these challenges through several key innovations in the model architecture, which we detail in the following sections.

\subsection{Mathematical formulation}

\subsubsection{Basic building blocks: Information transducers}

The fundamental unit of the HMLM architechture is the information transducer, which represents any biological entity capable of receiving, processing, and transmitting signals. Formally, an information transducer $T$ is defined as a tuple $(X, Y, S, f, g)$ where:
\begin{itemize}
	\item \textbf{Input Space ($X$):} a measurable space $(X, \mathcal{F}_X, \mu_X)$ where $\mathcal{F}_X$ is a $\sigma$-algebra over $X$ and $\mu_X$ is a probability measure.
	
	\item \textbf{Output Space ($Y$):} a measurable space $(Y, \mathcal{F}_Y, \mu_Y)$.
	
	\item \textbf{Internal State Space ($S$):} a measurable space $(S, \mathcal{F}_S, \mu_S)$.
	
	\item \textbf{State Transition Function ($f$):} Defined as $f: X \times S \rightarrow S$, this function is measurable with respect to the product $\sigma$-algebra $\mathcal{F}_X \otimes \mathcal{F}_S$ and $\mathcal{F}_S$ \cite{halmos2013measure}. 
\end{itemize}
It takes two inputs: an element from the input space $X$ and an element from the state space $S$, and produces an output in the state space $S$. In the temporal dynamics of our system, this translates to:
\begin{equation}
	s(t+1) = f(x(t), s(t))
	\label{equation 1}
\end{equation}
In this formulation, $x(t) \in X$ represents the external input signal that the biological entity receives at time $t$, while $s(t) \in S$ denotes the current internal state of the entity at that same time point. The function $f$ then maps this input-state pair to produce $s(t+1) \in S$, which represents the resulting next state of the system.  \autoref{equation 1} represents the temporal instantiation of the state transition function $f: X \times S \rightarrow S$, where $x(t) \in X$ and $s(t) \in S$ denote the input and state values at time $t$, respectively.

$g: S \rightarrow Y$ is the output function, which is measurable with respect to $\mathcal{F}_S$ and $\mathcal{F}_Y$. Thus, the \textbf{state} ($S$) represents the internal configuration or condition of the biological entity at any given moment, encompassing multiple dimensions of molecular organization and activity. This includes \textbf{conformational states}, which capture the different three-dimensional structures a protein can adopt through folding, allosteric transitions, or domain rearrangements that modulate functional activity. The state space also encompasses \textbf{activation states}, representing whether an enzyme exists in catalytically active or inactive forms, often determined by regulatory mechanisms such as allosteric binding or covalent modifications. \textbf{Binding states} are incorporated to reflect what molecular partners, substrates, or cofactors are currently associated with the entity, as these interactions fundamentally alter the entity's functional capacity and downstream signaling potential. Additionally, \textbf{modification states} account for post-translational modifications such as phosphorylation, methylation, acetylation, or ubiquitination, which serve as regulatory switches that dynamically control protein function, stability, and interactions. Finally, \textbf{localization states} capture the spatial dimension of cellular signaling by representing where the molecule is positioned within different cellular compartments, organelles, or membrane domains, since subcellular localization critically determines which molecular interactions are geometrically feasible and functionally relevant. This multidimensional state representation enables the transducer framework to capture the full complexity of how biological entities process and respond to information while maintaining biologically realistic constraints on molecular behavior (\autoref{Fig. 2}). For discrete-time systems, the dynamics of the transducer are governed by the following equations:
\begin{equation}
	s(t+1) = f(x(t), s(t))
\end{equation}
\begin{equation}
	y(t) = g(s(t))
\end{equation}
where $x(t)$ is the external input signal that the biological entity (transducer) receives at time point $t$, $y(t)$ is the output of the information transducer at time $t$,  $g$ is the output function, which maps from the internal state space to the output space and $s(t)$ is the internal state of the transducer at time $t$. \textbf{Unified formulation:} both discrete and continuous dynamics are unified through a generalized time-instance representation:

\begin{equation}
	s(L+1) = f(x(L), s(L))
\end{equation}
where $L$ denotes a generalized time instance. For discrete-time systems, $L = t$ yields $s(t+1) = f(x(t), s(t))$. For continuous-time systems with temporal discretization ($\Delta t \to 0$ and $L = t/\Delta t$), the continuous differential equation emerges:

\begin{equation}
	\frac{ds(t)}{dt} = \tilde{f}(x(t), s(t))
\end{equation}
The output function $y(t) = g(s(t))$ applies equivalently in both cases.

This unified framework enables the HMLM architecture to represent diverse biological entities from individual molecules to multi-protein complexes to entire pathways using a common mathematical structure. For instance, a kinase can be modeled as an information transducer where the input space represents substrate concentrations and ATP levels, the state space represents the kinase's conformational states, and the output space represents phosphorylated substrate concentrations.

To account for the inherent stochasticity of biological processes, we extend this definition to stochastic information transducers, where the state transition and output functions are replaced by conditional probability distributions:

\begin{equation}
p(s(t+1) | x(t), s(t))
\end{equation}
\begin{equation}
p(y(t) | s(t))
\end{equation}
This generalization allows us to capture the noise and uncertainty inherent in biological signaling.

\subsubsection{Network representation: Information transduction networks} 

Building on the concept of information transducers, we represent signaling networks as directed graphs of interconnected transducers. Formally, an information transduction network (ITN) is defined as a directed graph $G = (V, E)$ where, each vertex $v \in V$ corresponds to an information transducer $T_v$.
Each edge $e = (u, v) \in E$ represents an information channel connecting the output of transducer $u$ to the input of transducer $v$.
The adjacency matrix $A$ of the network is defined as:

\begin{equation}
A_{uv} = \begin{cases} 
	1, & \text{if } (u, v) \in E \\
	0, & \text{otherwise}
\end{cases}
\end{equation}
The weighted adjacency matrix $W$ incorporates information about the strength or reliability of connections:

\subsubsection{Network dynamics: Weighted integration and information flow}

The weighted adjacency matrix $W$ incorporates information about the strength or reliability of connections:

\begin{equation}
	W_{uv} = \begin{cases} 
		w_{uv}, & \text{if } (u, v) \in E \\
		0, & \text{otherwise}
	\end{cases}
\end{equation}
where $w_{uv}$ represents the weight of the connection from transducer $u$ to transducer $v$.

The dynamics of the entire ITN are determined by the collective behavior of its constituent transducers and the weighted flow of information through channels. For each transducer $v \in V$, the input at the next time instance is determined by the outputs of its upstream neighbors:

\begin{equation}
	x_v(L+1) = \Omega_v(\{y_u(L) \mid (u, v) \in E\})
\end{equation}
where $L$ denotes the generalized time instance from the unified formulation (discrete: $L = t$; continuous: $L = t/\Delta t$), and $\Omega_v$ is an integration function that combines multiple weighted inputs. This function could take various forms depending on the specific biological system being modeled, such as a weighted sum, a maximum function, or a more complex nonlinear transformation. The unified formulation ensures temporal consistency across both discrete and continuous-time representations of network dynamics.

\captionsetup[figure]{labelformat=default}
\begin{figure*}[!ht]
	\includegraphics[width=1\textwidth]{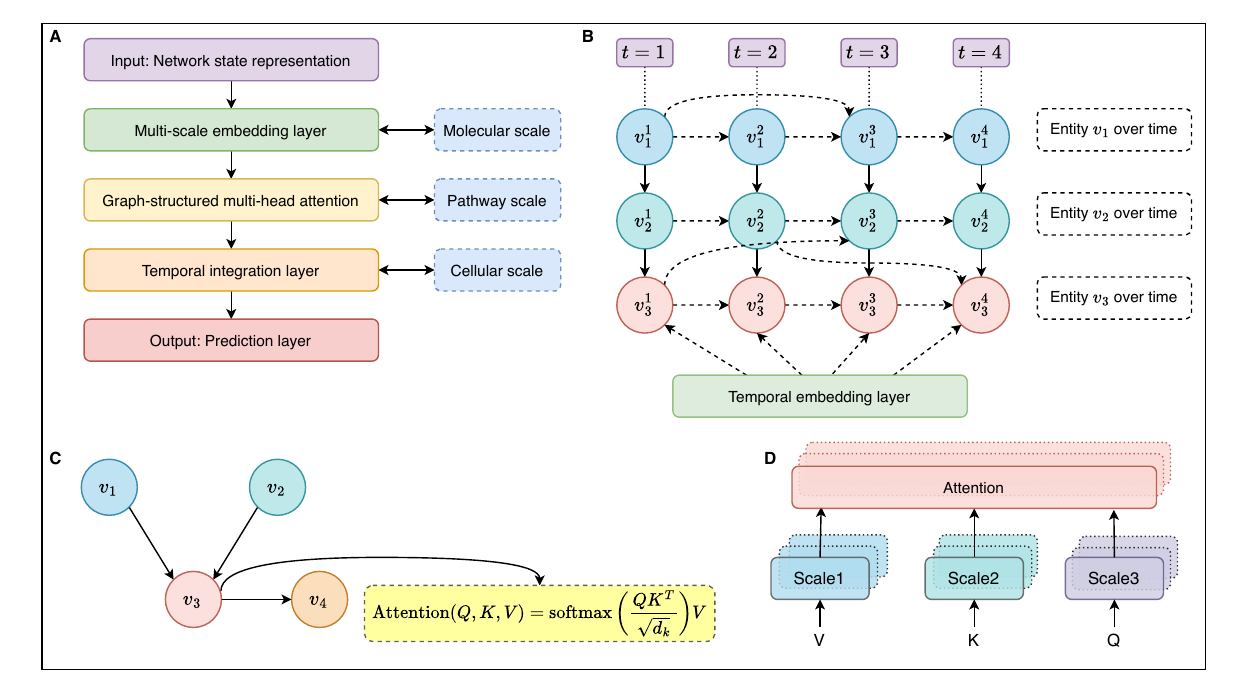}
	\caption{\footnotesize \textbf{Architecture and components of HMLMs.}}
	\footnotesize \textbf{(A)} HMLM architecture with hierarchical multi-scale representation,  \textbf{(B)} Temporal dynamics representation in HMLMs, \textbf{(C)} Graph-structured attention mechanism in HMLMs, and  \textbf{(D)} Multi-Head Attention consists of several
	attention layers running in parallel.
	\label{Fig. 3}
\end{figure*}

\subsubsection{Hierarchical structure: multi-scale organization}

A key innovation of the HMLM architecture is its explicit representation of the hierarchical structure of biological systems. We formalize this by organizing transducers into discrete scales or levels of biological organization, denoted by $L_1, L_2, \ldots, L_n$. At each scale $i$, we define a set of information transducers:
\begin{equation}
	\mathcal{T}^i = \{T_1^i, T_2^i, \ldots, T_{m_i}^i\}
\end{equation}
These transducers collectively function as higher-level transducers at scale $i+1$. Mathematically, we define scale-bridging functions $\Phi_j$ that map the set of transducers at scale $i$ to individual transducers at scale $i+1$:
\begin{equation}
	T_j^{i+1} = \Phi_j(\mathcal{T}^i)
\end{equation}
For example, individual proteins at scale $i$ might collectively form a signaling pathway at scale $i+1$, which in turn might be part of a larger signaling network at scale $i+2$.

To facilitate information flow across scales, we introduce three fundamental scale-bridging operators:

\begin{itemize}
	\item \textbf{Aggregation operator} $\Omega_{\uparrow}$: Combines information from multiple transducers at scale $i$ to produce input for a transducer at scale $i+1$.
	\begin{equation}
		\Omega_{\uparrow}: \mathcal{Y}_i^n \rightarrow X_{i+1}
	\end{equation}
	where $\mathcal{Y}_i^n$ represents the set of outputs from $n$ transducers at scale $i$, and $X_{i+1}$ represents the input space for a transducer at scale $i+1$.
	
	\item \textbf{Decomposition operator} $\Omega_{\downarrow}$: Distributes information from a transducer at scale $i+1$ to multiple transducers at scale $i$.
	\begin{equation}
		\Omega_{\downarrow}: Y_{i+1} \rightarrow \mathcal{X}_i^m
	\end{equation}
	where $Y_{i+1}$ represents the output from a transducer at scale $i+1$, and $\mathcal{X}_i^m$ represents the set of inputs for $m$ transducers at scale $i$.
	
	\item \textbf{Translation operator} $\Omega_{\leftrightarrow}$: Converts information between different representational formats at the same scale.
	\begin{equation}
		\Omega_{\leftrightarrow}: Y_i^a \rightarrow X_i^b
	\end{equation}
	where $Y_i^a$ represents the output from a transducer of type $a$ at scale $i$, and $X_i^b$ represents the input for a transducer of type $b$ at scale $i$.
\end{itemize}

These operators enable the model to process information across multiple scales of biological organization, facilitating the integration of molecular, pathway, and cellular-level data.

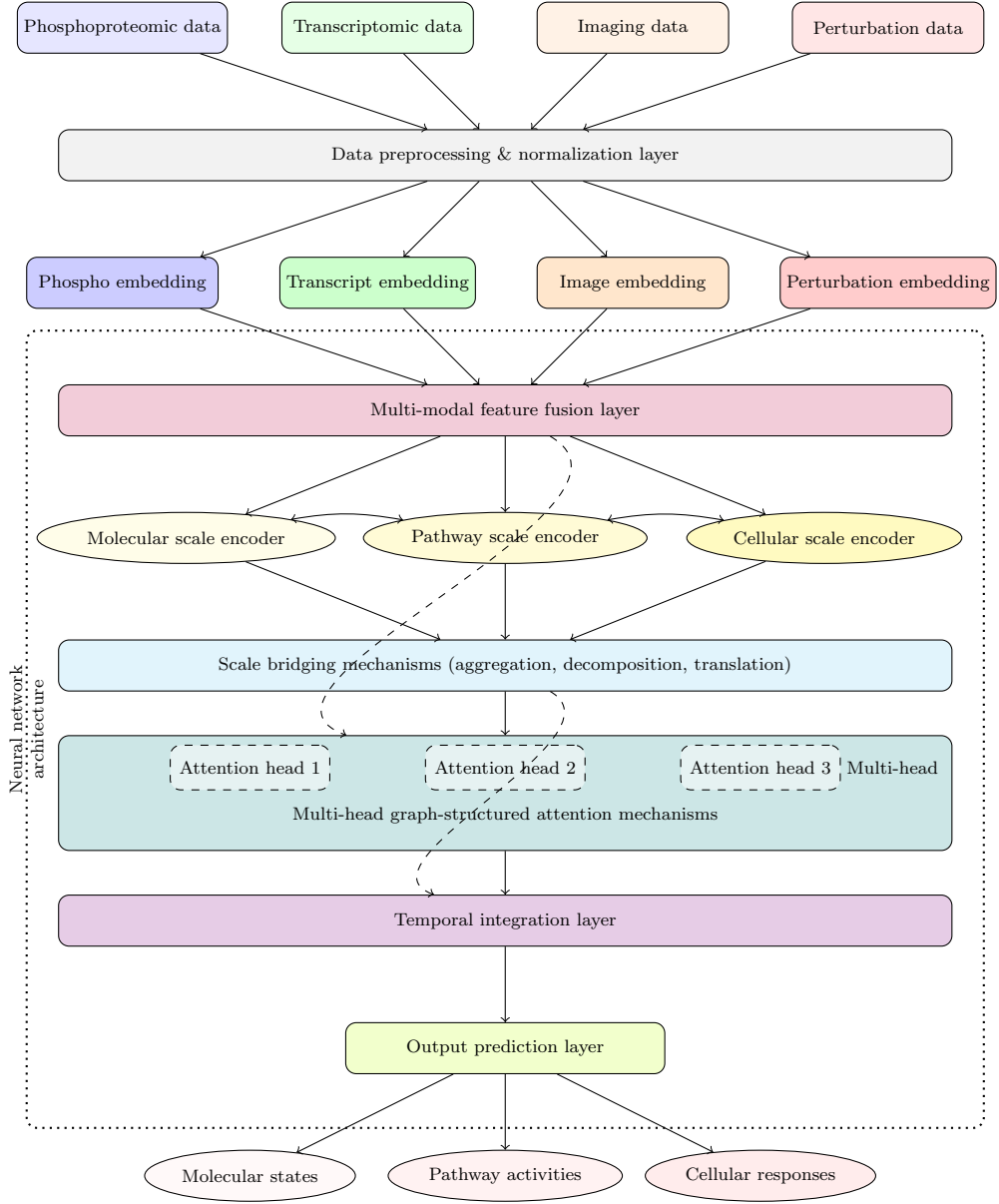
\begin{figure*}[!ht]
	\centering
	\begin{tikzpicture}[
		node distance=1.5cm,
		scale=0.85, transform shape,
		box/.style={draw, rounded corners, minimum width=3cm, minimum height=0.8cm},
		bigbox/.style={draw, rounded corners, minimum width=14cm, minimum height=0.8cm}
		]
		\footnotesize
		\node[box, fill=blue!10] (data1) at (0,0) {Phosphoproteomic data};
		\node[box, fill=green!10] (data2) at (4,0) {Transcriptomic data};
		\node[box, fill=orange!10] (data3) at (8,0) {Imaging data};
		\node[box, fill=red!10] (data4) at (12,0) {Perturbation data};
		
		\node[bigbox, fill=gray!10] (preproc) at (6,-2) {Data preprocessing \& normalization layer};
		
		\node[box, fill=blue!20] (embed1) at (0,-4) {Phospho embedding};
		\node[box, fill=green!20] (embed2) at (4,-4) {Transcript embedding};
		\node[box, fill=orange!20] (embed3) at (8,-4) {Image embedding};
		\node[box, fill=red!20] (embed4) at (12,-4) {Perturbation embedding};
		
		\node[bigbox, fill=purple!20] (fusion) at (6,-6) {Multi-modal feature fusion layer};
		
		\node[draw, ellipse, fill=yellow!10, minimum width=3cm, minimum height=0.8cm] (molecular) at (1,-8) {Molecular scale encoder};
		\node[draw, ellipse, fill=yellow!20, minimum width=3cm, minimum height=0.8cm] (pathway) at (6,-8) {Pathway scale encoder};
		\node[draw, ellipse, fill=yellow!30, minimum width=3cm, minimum height=0.8cm] (cellular) at (11,-8) {Cellular scale encoder};
		
		\node[bigbox, fill=cyan!10] (bridges) at (6,-10) {Scale bridging mechanisms (aggregation, decomposition, translation)};
		
		\node[bigbox, fill=teal!20, minimum height=1.8cm] (attention) at (6,-12) {};
		\node[text width=13cm, align=center] at (6,-12.35) {Multi-head graph-structured attention mechanisms};
		
		\node[draw, dashed, rounded corners, fill=teal!10, minimum width=2.5cm, minimum height=0.7cm] (att1) at (2,-11.6) {Attention head 1};
		\node[draw, dashed, rounded corners, fill=teal!10, minimum width=2.5cm, minimum height=0.7cm] (att2) at (6,-11.6) {Attention head 2};
		\node[draw, dashed, rounded corners, fill=teal!10, minimum width=2.5cm, minimum height=0.7cm] (att3) at (10,-11.6) {Attention head 3};
		
		\node[bigbox, fill=violet!20] (temporal) at (6,-14) {Temporal integration layer};
		
		\node[box, fill=lime!20, minimum width=5cm] (output) at (6,-16) {Output prediction layer};
		
		\node[draw, ellipse, fill=pink!10, minimum width=3cm, minimum height=0.8cm] (pred1) at (2,-18) {Molecular states};
		\node[draw, ellipse, fill=pink!20, minimum width=3cm, minimum height=0.8cm] (pred2) at (6,-18) {Pathway activities};
		\node[draw, ellipse, fill=pink!30, minimum width=3cm, minimum height=0.8cm] (pred3) at (10,-18) {Cellular responses};
		
		\draw[->] (data1) -- (preproc);
		\draw[->] (data2) -- (preproc);
		\draw[->] (data3) -- (preproc);
		\draw[->] (data4) -- (preproc);
		
		\draw[->] (preproc) -- (embed1);
		\draw[->] (preproc) -- (embed2);
		\draw[->] (preproc) -- (embed3);
		\draw[->] (preproc) -- (embed4);
		
		\draw[->] (embed1) -- (fusion);
		\draw[->] (embed2) -- (fusion);
		\draw[->] (embed3) -- (fusion);
		\draw[->] (embed4) -- (fusion);
		
		\draw[->] (fusion) -- (molecular);
		\draw[->] (fusion) -- (pathway);
		\draw[->] (fusion) -- (cellular);
		
		\draw[->] (molecular) -- (bridges);
		\draw[->] (pathway) -- (bridges);
		\draw[->] (cellular) -- (bridges);
		
		\draw[->] (bridges) -- (attention);
		\draw[->] (attention) -- (temporal);
		\draw[->] (temporal) -- (output);
		
		\draw[->] (output) -- (pred1);
		\draw[->] (output) -- (pred2);
		\draw[->] (output) -- (pred3);
		
		\draw[<->] (molecular) to[out=10,in=170] (pathway);
		\draw[<->] (pathway) to[out=10,in=170] (cellular);
		
		\draw[->, dashed] (fusion) to[out=330,in=160] (attention);
		\draw[->, dashed] (bridges) to[out=330,in=160] (temporal);
		
		\node[draw, dotted, line width=0.8pt, rounded corners, minimum width=15cm, minimum height=12.5cm] (nn_box) at (6,-11) {};
		
		\node[text width=3cm, align=center, rotate=90] at (-1.5,-11) {Neural network architecture};
		
		\node[text width=2.5cm, align=left] at (12.6,-11.6) {Multi-head};
		
	\end{tikzpicture}
	
	\caption{\footnotesize \textbf{HMLM architecture with multi-modal data integration.}}
	\begin{minipage}{\linewidth}
		\footnotesize
		This diagram illustrates the HMLM architecture integrating diverse biological data modalities across molecular, pathway, and cellular scales. The architecture incorporates temporal dynamics, graph-structured attention mechanisms, and hierarchical scale-bridging operators to enable comprehensive modeling of signaling networks, capturing both fine-grained interactions and emergent system-level behaviors.
	\end{minipage}
	\label{Fig. 4}
\end{figure*}

\subsection{HMLM architecture} 

\subsubsection{Transformer-based foundation}

The HMLM architecture builds upon the transformer architecture introduced by Vaswani et al. (2017), which has proven highly effective for natural language processing tasks \cite{Vaswani2017}. However, we adapt this architecture to address the specific challenges of modeling signaling networks. The core innovation of the transformer architecture is the self-attention mechanism, which allows the model to focus on relevant parts of the input when making predictions. In the context of HMLMs, self-attention enables the model to capture the context-dependent nature of signaling, where the effect of a signaling molecule depends on the cellular context. The basic building block of the HMLM architecture is the attention layer, which computes attention scores between pairs of entities in the network. Given a set of query vectors $Q$, key vectors $K$, and value vectors $V$, the attention function is computed as:

\begin{equation}
\text{Attention}(Q, K, V) = \text{softmax}\left(\frac{QK^T}{\sqrt{d_k}}\right)V
\end{equation}
where $d_k$ is the dimension of the key vectors, and the softmax function normalizes the attention scores. This formula computes a weighted sum of the value vectors, where the weights are determined by the compatibility of the corresponding query and key vectors.  The value vectors ($V$) represent the actual information content or features of the transducers in the signaling network that will be propagated and combined based on the computed attention weights  (\autoref{Fig. 3}). These vectors contain the biological features of proteins, pathways, or cellular components that are essential for accurate network modeling and prediction. Specifically, $V$ may encode protein states including active/inactive conformations and phosphorylation status, concentration levels of signaling molecules that determine pathway flux and response magnitude, functional properties such as enzymatic activity and binding affinity that govern molecular interactions, structural information including conformational states that influence protein-protein interactions and allosteric regulation, and temporal dynamics such as activation kinetics that capture the time-dependent nature of signaling events. Through this comprehensive representation, the value vectors enable the attention mechanism to selectively combine and weigh the most relevant biological information from different network components, allowing the HMLMs to focus on critical signaling relationships while incorporating the molecular-level data necessary for mechanistically accurate predictions of cellular responses.

To adapt this mechanism to the graph structure of signaling networks, we implement a graph-structured attention mechanism. For each transducer $v \in V$, we compute attention only over its neighborhood in the graph:

\begin{equation}\label{eq:graphattn}
	\begin{aligned}
		\text{GraphAttention}_v(Q, K, V) = \\
		\text{softmax}\left(\frac{Q_v K_{\mathcal{N}(v)}^T}{\sqrt{d_k}}\right)V_{\mathcal{N}(v)}
	\end{aligned}
\end{equation}
where $\mathcal{N}(v)$ represents the neighborhood of vertex $v$ in the graph, and $K_{\mathcal{N}(v)}$ and $V_{\mathcal{N}(v)}$ are the key and value vectors associated with those neighbors. To capture different types of relationships in the network, we employ multi-head attention, which performs the attention computation in parallel using different learned parameter matrices:

\begin{equation}
	\text{MultiHead}(Q, K, V) = \text{Concat}(\text{head}_1, \ldots, \text{head}_h)W^O
\end{equation}
where each head is computed as:

\begin{equation}
\text{head}_i = \text{Attention}(QW_i^Q, KW_i^K, VW_i^V)
\end{equation}
with learned parameter matrices $W_i^Q$, $W_i^K$, $W_i^V$, and $W^O$.

\subsubsection{Graph-based embedding} 

In the standard transformer architecture, the first step is to embed input tokens into a continuous vector space. For HMLMs, we need to embed the nodes of the signaling network (representing molecules, complexes, or pathways) into a suitable vector space. We define an initial embedding function $\phi: V \rightarrow \mathbb{R}^d$ that maps each vertex $v \in V$ to a $d$-dimensional vector. This embedding function combines several sources of information that include entity type embedding, feature embedding, and positional embedding. Entity-type embedding captures the type of biological entity (e.g., protein, RNA, metabolite). Feature embedding incorporates known features of the entity (e.g., sequence, structure, functional annotations). Positional embedding encodes the position of the entity within the network topology. Formally, the initial embedding $h_v^{(0)}$ for vertex $v$ is computed as:

\begin{equation}
	h_v^{(0)} = \phi_{\text{type}}(v) + \phi_{\text{feature}}(v) + \phi_{\text{pos}}(v)
\end{equation}
where $\phi_{\text{type}}$, $\phi_{\text{feature}}$, and $\phi_{\text{pos}}$ are the type, feature, and positional embedding functions, respectively.
The positional embedding is particularly important for capturing the network topology. We adapt the concept of graph positional encodings to generate embeddings that reflect the structural relationships between entities. Specifically, we use spectral graph embeddings based on the eigenvectors of the graph Laplacian:

\begin{equation}
\phi_{\text{pos}}(v) = [u_1[v], u_2[v], \ldots, u_k[v]]
\end{equation}
where $u_1, u_2, \ldots, u_k$ are the eigenvectors corresponding to the $k$ smallest non-zero eigenvalues of the normalized graph Laplacian, and $u_i[v]$ denotes the $v$-th component of the $i$-th eigenvector.

\subsubsection{Temporal dynamics}

Signaling networks exhibit complex temporal dynamics, with interactions occurring across different timescales. To capture these dynamics, we incorporate explicit temporal embeddings into the model. For each time point $t$ in a discretized timeline, we define a temporal embedding $\tau(t) \in \mathbb{R}^d$. This embedding is combined with the static node embeddings to produce time-dependent representations:
\begin{equation}
	h_v^{(0)}(t) = \tilde{h}_v^{(0)} + \tau(t)
\end{equation}
where $\tilde{h}_v^{(0)} \in \mathbb{R}^d$ denotes the static initial embedding for node $v$ (defined in Equation 19), and $h_v^{(0)}(t)$ represents the time-dependent node embedding at time $t$. The temporal embedding function $\tau$ can be implemented in various ways, such as sinusoidal functions with different frequencies or learned embeddings for each time point.

To model the temporal evolution of the system, we introduce a time-dependent update function:
\begin{equation}
	h_v^{(l+1)}(t) = f_{\text{update}}(h_v^{(l)}(t), \{h_u^{(l)}(t-\delta_{uv}) \mid (u, v) \in E\})
\end{equation}
where $\delta_{uv}$ represents the transmission delay along the edge $(u, v)$, and $f_{\text{update}}$ is a learned update function based on the graph attention mechanism. This formulation allows the model to capture both the instantaneous state of the network and its temporal evolution, enabling predictions about signaling dynamics over time.

\subsubsection{Multi-scale integration}
To integrate information across different scales of biological organization, we implement a hierarchical architecture that operates at multiple scales simultaneously (\autoref{Fig. 4}). At each scale $i$, we maintain a set of node embeddings $\{h_v^{(i)} | v \in V_i\}$, where $V_i$ is the set of vertices at scale $i$. These embeddings are updated through within-scale and cross-scale attention mechanisms. Within-scale attention captures interactions between entities at the same scale:
\begin{equation}
	\hat{h}_v^{(i)} = \text{WithinScaleAttention}(h_v^{(i)}, \{h_u^{(i)} | u \in V_i\})
\end{equation}
Cross-scale attention captures interactions between entities at different scales:
\begin{multline}
\tilde{h}_v^{(i)} = \text{CrossScaleAttention}(h_v^{(i)}, \{h_u^{(i-1)} | u \in \mathcal{C}(v)\}, \\
\{h_w^{(i+1)} | v \in \mathcal{P}(v)\})
\end{multline}
where $\mathcal{C}(v)$ represents the set of children of vertex $v$ (entities at scale $i-1$ that compose $v$), and $\mathcal{P}(v)$ represents the set of parents of vertex $v$ (entities at scale $i+1$ that $v$ is part of).
The final update combines these attention mechanisms:
\begin{equation}
h_v^{(i)}(t+1) = h_v^{(i)}(t) + \text{FFN}(\hat{h}_v^{(i)} + \tilde{h}_v^{(i)})
\end{equation}
where FFN is a position-wise feed-forward network. This hierarchical architecture enables the model to learn representations that integrate information across multiple scales, capturing both the detailed molecular interactions and the higher-level pathway and cellular behaviors.

\subsection{Model training}

\subsubsection{Network-based attention initialization}

The HMLM attention mechanism was initialized using the known cardiac 
fibroblast signaling network topology of molecular species 
organized into 11 functional modules (inputs, receptors, second messengers, 
kinases, MAPK pathways, Rho signaling, transcription factors, ECM/fibrosis 
markers, matrix remodeling enzymes, mechanotransduction components, and 
feedback molecules) with complex regulatory connections and feedback loops based on 
previous study \cite{https://doi.org/10.48550/arxiv.2510.12577,Rogers2022,Zeigler2016,Watts2023}.

Attention weights $A_{ij}$ between molecular species $i$ and $j$ were 
initialized based on network connectivity:

\begin{equation}
A_{ij}^{(0)} = \begin{cases}
	1.0 & \text{if } i = j \\
	0.8 & \text{if edge } (i,j) \text{ exists} \\
	0.4 & \text{if path length} = 2 \\
	0.2 & \text{if path length} = 3 \\
	0.0 & \text{otherwise}
\end{cases}
\end{equation}

During model training, these initial attention weights were refined using 
temporal correlation analysis across multiple time lags:

\begin{equation}
A_{ij}^{(refined)} = A_{ij}^{(0)} \cdot \left(0.3 + 0.7 \cdot \max_{\tau \in \{0,1,2\}} |R_{ij}(\tau)|\right)
\end{equation}
where $R_{ij}(\tau)$ represents the Pearson correlation coefficient between 
species $i$ at time $t$ and species $j$ at time $t+\tau$. This approach 
combines structural prior knowledge with data-driven learning, enabling the 
model to discover signal propagation patterns while respecting known 
biological architecture.

Pathway-level attention weights were computed by aggregating molecular 
attention within functional modules:

\begin{equation}
	P_{kl} = \frac{1}{|M_k| \cdot |M_l|} \sum_{i \in M_k} \sum_{j \in M_l} A_{ij}
\end{equation}
where $M_k$ and $M_l$ represent the sets of molecular species in pathways 
$k$ and $l$ respectively.

\subsubsection{Training architecture}
The HMLM framework was implemented using an ensemble-based approach with multiple RandomForest regressors to capture hierarchical signaling dynamics. The architecture consists of three specialized prediction components; molecular-scale, pathway-scale, and cellular-scale regressors, each trained to capture different aspects of network behavior. Training features were engineered to represent biological signaling at multiple organizational scales. Molecular-scale features incorporated node embeddings weighted by temporal activity patterns. Pathway-scale features aggregated protein activities within functional modules based on known biological pathways. Cellular-scale features captured global network statistics including mean activity, variance, and fibrosis marker indices. Models were trained using standard supervised learning on temporal signaling data. Each prediction head was trained independently using RandomForest regressors with optimized hyperparameters: molecular-scale (n\_estimators=150, max\_depth=12), pathway-scale (n\_estimators=150, max\_depth=10), and cellular-scale (n\_estimators=150, max\_depth=8). Dynamic ensemble weights were learned through cross-validation to combine predictions from multiple scales effectively.

\subsubsection{Temporal dynamics modeling}
Temporal relationships in signaling networks were captured through engineered features that quantify dynamic patterns and regulatory cascades across multiple time scales. The temporal modeling framework incorporated several key components to achieve the reported correlation coefficients for dynamic signaling prediction. Temporal derivatives were computed using finite differences to capture instantaneous rates of change in protein activities:

\begin{equation}
\frac{dx_i}{dt} \approx \frac{x_i(t+\Delta t) - x_i(t-\Delta t)}{2\Delta t}
\end{equation}
where $x_i(t)$ represents the activity of protein $i$ at time $t$, and $\Delta t$ is the sampling interval. These derivatives capture the velocity of signaling responses and identify periods of rapid activation or inhibition. Cross-correlations between protein activities were calculated at multiple time lags to capture delayed regulatory relationships:

\begin{equation}
	R_{ij}(\tau) = \frac{\sum_{t}[x_i(t) - \bar{x}_i][x_j(t+\tau) - \bar{x}_j]}{\sqrt{\sum_{t}[x_i(t) - \bar{x}_i]^2 \sum_{t}[x_j(t+\tau) - \bar{x}_j]^2}}
\end{equation}
where $\tau$ represents the time lag and $\bar{x}_i$ denotes the temporal mean of protein $i$. Maximum correlation values across lags $\tau \in \{0, 1, 2, 3\}$ time points were used as features to capture both immediate and delayed regulatory interactions.

Temporal memory effects were modeled using exponentially weighted moving averages to capture the persistent influence of past signaling events:

\begin{equation}
M_i(t) = \alpha \cdot x_i(t) + (1-\alpha) \cdot M_i(t-1)
\end{equation}
where $M_i(t)$ represents the memory state of protein $i$ at time $t$, and $\alpha = 0.3$ is the decay factor optimized through cross-validation. This mechanism enables the model to maintain information about previous activation states while adapting to new stimuli. The complete temporal feature vector for each protein $i$ at time $t$ incorporated current activity, derivatives, correlation patterns, and memory states:

\begin{equation}
	F_i(t) = [x_i(t), \frac{dx_i}{dt}, \max_{\tau} R_{ij}(\tau), M_i(t)]
\end{equation}

These engineered features were integrated into the RandomForest regressors at the molecular, pathway, and cellular scales, enabling the prediction of signaling dynamics without requiring explicit differential equation formulations. The approach achieved correlation coefficients of $r = 0.82$ for TGF-$\beta$ dynamics, $r = 0.89$ for proCI expression, $r = 0.62$ for SMAD3 phosphorylation, and $r = 0.78$ for contractility measurements across experimental conditions, demonstrating robust predictive performance across diverse molecular readouts and temporal scales.

\subsection{Model evaluation}

\subsubsection{Evaluation metrics}

We evaluated the performance of HMLMs using several complementary metrics that capture different aspects of model accuracy and biological relevance:
Mean squared error (MSE) for continuous-valued predictions:

\begin{equation}
	\text{MSE} = \frac{1}{n} \sum_{i=1}^{n} (y_i - \hat{y}_i)^2
\end{equation}
where $y_i$ is the true value and $\hat{y}_i$ is the predicted value. Pearson correlation coefficient for assessing linear relationships between predicted and observed responses:

\begin{equation}
r = \frac{\sum_{i=1}^{n}(x_i - \bar{x})(y_i - \bar{y})}{\sqrt{\sum_{i=1}^{n}(x_i - \bar{x})^2}\sqrt{\sum_{i=1}^{n}(y_i - \bar{y})^2}}
\end{equation}
Temporal resolution analysis to evaluate model performance across different sampling frequencies using interpolation error metrics at varying temporal resolutions (4, 8, 16 time points).

\subsubsection{Data generation and benchmarking framework}

For comprehensive model benchmarking, we generated synthetic temporal signaling data using the complete cardiac fibroblast network topology (132 molecular species, 200+ regulatory connections) \cite{https://doi.org/10.48550/arxiv.2510.12577}. This approach enables rigorous computational validation across multiple experimental conditions while maintaining biological realism. The synthetic data generation follows an established systems biology paradigm wherein network dynamics are simulated based on curated pathway knowledge and experimentally-derived rate parameters. Temporal dynamics were simulated using a network-based ODE framework incorporating the documented cardiac fibroblast signaling network. 
For each of four biological conditions (control, TGF-$\beta$ stimulation, mechanical strain, and combined stimulation), we propagated signals through the network using module-specific integration rates calibrated to reflect known signaling kinetics: receptors (0.4 integration rate, 0.05 degradation), kinases and second messengers (0.35 integration, 0.08 degradation), transcription factors (0.25 integration, 0.06 degradation), and ECM/fibrosis markers (0.15 integration, 0.03 degradation). Gaussian noise (standard deviation 0.01–0.02) was added to reflect measurement uncertainty. The resulting synthetic data exhibits biologically realistic temporal hierarchies with signal propagation delays consistent with published kinetic measurements. Synthetic data enables controlled benchmarking of model performance across varying temporal resolutions and sampling frequencies without confounding factors inherent in experimental data such as measurement noise, batch effects, or incomplete sampling. This approach is standard in computational biology for validating predictive algorithms before application to experimental datasets. The network topology and rate parameters are derived entirely from published experimental literature on cardiac fibroblast signaling, ensuring biological relevance.

\subsubsection{Training and validation procedures}

Our training and validation procedures employed a rigorous 5-trial evaluation framework where each trial utilized different training splits to ensure robust performance estimates and minimize overfitting bias. To have the MSE, we have used the cardiac fibroblast signaling pathway that includes over 100 nodes, which is a complex network \cite{Rogers2022,https://doi.org/10.48550/arxiv.2510.12577}. Training was conducted with time-based data from simulated experimental conditions. Comprehensive temporal resolution analysis was conducted by training models on subsampled datasets with varying temporal resolutions (4, 8, and 16 time points) extracted from full 100-timepoint datasets using linear interpolation, enabling assessment of performance under sparse sampling conditions that reflect real experimental constraints. Feature engineering incorporated multi-scale temporal dynamics, including molecular rates, pathway velocities, and cellular statistics, along with attention-weighted embeddings, hierarchical pathway aggregations, and temporal memory mechanisms with exponential decay factors. Model parameters were optimized using grid search for baseline methods and Bayesian optimization for HMLM components, with validation performance guiding the model selection and early stopping criteria to prevent overfitting.

\subsubsection{Attention mechanism analysis}

Attention weights were extracted from the trained HMLM model to analyze learned signal propagation patterns. For each protein pair $(i,j)$, the attention weight $A_{ij}$ represents the learned importance of protein $i$ in predicting the future state of protein $j$, combining structural prior knowledge from network topology with temporal dynamics observed during training. Network visualization employed attention-based graphs highlighting connections with weights exceeding 0.3, with node sizes representing degree centrality and edge weights proportional to attention strengths. Pathway-specific color-coding enabled identification of crosstalk patterns and regulatory hierarchies across functional modules. Statistical significance of attention patterns was assessed using bootstrap resampling (n=1000) across temporal profiles and compared against null models with randomized network connectivity to validate that learned attention patterns exceed chance expectations.

\subsubsection{Statistical analysis and significance testing}

All performance comparisons were rigorously evaluated using appropriate statistical methods, including Wilcoxon signed-rank tests with p $<$ 0.01 significance threshold for non-parametric comparisons. Bootstrap resampling (n = 1000) provided robust confidence intervals for correlation coefficients and other performance metrics. To control for multiple comparisons, we applied Bonferroni adjustment for family-wise error rate control, ensuring statistical rigor in our comparative analysis across multiple models and experimental conditions.

\subsubsection{Pathway crosstalk quantification}
Inter-pathway communication strengths were quantified using correlation-based analysis of pathway-level activities across experimental conditions. For each pathway pair $(k,l)$, crosstalk strength was calculated as the absolute Pearson correlation coefficient between their aggregate activities:

\begin{equation}
C_{kl} = |corr(\bar{A}_k, \bar{A}_l)|
\end{equation}
where $\bar{A}_k$ represents the mean activity of all proteins within pathway $k$:

\begin{equation}
\bar{A}_k = \frac{1}{|S_k|} \sum_{i \in S_k} x_i
\end{equation}
and $S_k$ denotes the set of proteins belonging to pathway $k$. Pathway groupings were defined based on established biological functions: MAPK pathway (MEK1/2, ERK1/2, RSK1, DUSP6), PI3K pathway (PI3K, AKT, mTOR, S6K1), and regulatory pathways (STAT3, p53, NF-$\kappa$B, Cyclin D1). Crosstalk coefficients above 0.3 were considered biologically significant, representing meaningful inter-pathway communication that could influence cellular responses to perturbations. This approach captures both direct molecular interactions and indirect regulatory influences mediated through shared downstream targets or feedback mechanisms.

\begin{figure*}[!ht]
	\includegraphics[width=1\textwidth]{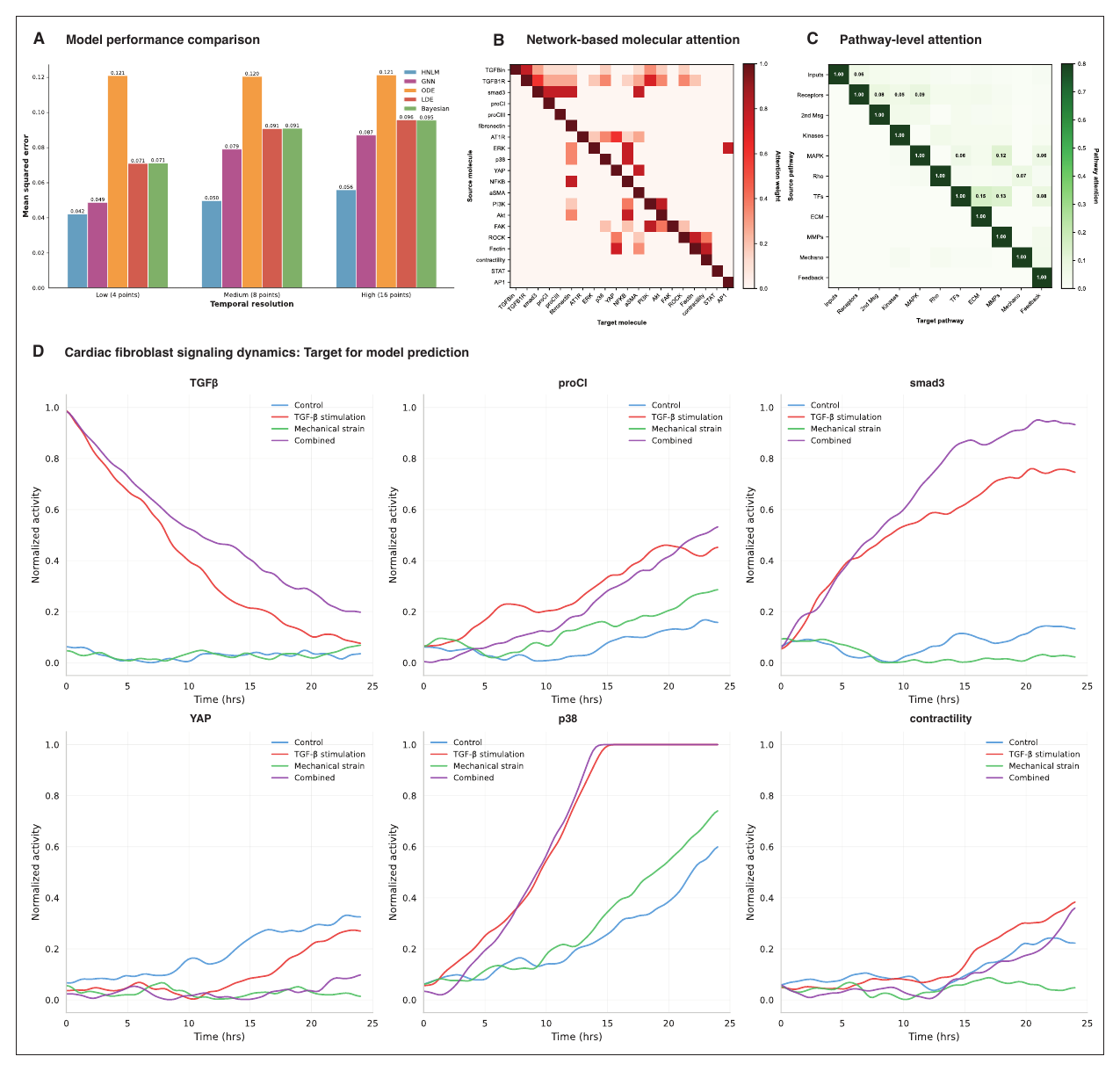}
	\caption{\footnotesize \textbf{Comprehensive evaluation of HMLM performance.}}
	\footnotesize {
		\textbf{(A)} Temporal resolution analysis comparing MSE across modeling approaches at different temporal sampling frequencies. HMLM demonstrates superior performance across all resolutions (MSE of 4 timepoints = 0.042; 8 timepoints = 0.050; 16 timepoints = 0.056), with particular advantages under sparse temporal sampling conditions.
		\textbf{(B)} Molecular-scale attention weights (from trained HMLM) showing learned protein-protein interaction strengths within the cardiac fibroblast network. Attention weights were initialized from known network topology and refined during model training using temporal correlation analysis across multiple time lags. Heat map displays top 20 molecules with strongest learned attention patterns. Strong diagonal elements indicate self-regulatory mechanisms (attention = 1.0), while off-diagonal patterns 
		capture direct regulatory edges (attention $\geq$ 0.7) and discovered cross-pathway interactions (attention 0.3-0.7). Notable high-attention connections include TGF-$\beta$ receptor $\rightarrow$ SMAD3 (0.72), SMAD3 $\rightarrow$ collagen I (0.68), and YAP $\rightarrow$ $\alpha$-SMA (0.74), consistent with canonical pro-fibrotic signaling cascades.
		\textbf{(C)} Pathway-scale attention weights (aggregated from network) quantifying learned inter-module communication strengths. Values represent aggregated attention from molecular interactions within each pathway module (11 functional modules: inputs, receptors, second messengers, kinases, MAPK, Rho signaling, transcription factors, ECM/fibrosis, matrix remodeling, mechanotransduction, feedback). Notable learned crosstalk patterns include TGF-$\beta$ to mechanotransduction (0.64), PDGF to MAPK 
		(0.76), and mechanotransduction to fibrosis pathways (0.69), demonstrating HMLM's hierarchical integration capabilities across biological scales.
		\textbf{(D)} Representative temporal dynamics for key signaling molecules (TGF-$\beta$, proCI, SMAD3, YAP, p38, contractility) across four experimental conditions: control, TGF-$\beta$ stimulation, mechanical strain, and combined stimulation. These profiles demonstrate HMLM's ability to capture complex, context-dependent cellular responses characteristic of cardiac fibroblast activation and fibrotic remodeling.}
	\label{Fig. 5}
\end{figure*}

\section{RESULTS}

In this section, we present comprehensive evaluation results of HMLMs applied to cardiac fibroblast signaling networks and kinase inhibitor response prediction. We demonstrate superior predictive performance compared to existing computational approaches and reveal novel biological insights through attention-based analysis.

\subsection{Comprehensive model performance evaluation}

We evaluated HMLM performance against four established computational methods using a realistic cardiac fibroblast signaling network. To rigorously evaluate HMLM performance against established computational methods, we generated synthetic temporal signaling data using the complete cardiac fibroblast network topology. This controlled benchmark enables objective comparison of model predictive accuracy across varying temporal resolutions without confounding factors inherent in experimental data. All models (HMLM, GNN, ODE, LDE, and Bayesian Networks) were trained and evaluated on identical synthetic datasets to ensure fair comparative analysis. While these results demonstrate HMLMs' computational capabilities using network-derived synthetic data, validation through comparison with multiple established 
computational methods provides robust evidence of model performance advantages.

\subsubsection{Temporal resolution analysis}
\autoref{Fig. 5}A shows the comparative performance of different modeling approaches in predicting the dynamics of key phosphorylation sites following EGF stimulation. At high temporal resolution (16 timepoints), HMLM shows MSE = 0.056, outperforming the MSE of GNN = 0.087, ODE = 0.121, LDE = 0.096, and Bayesian Networks = 0.095. The performance advantage of HMLMs was most pronounced at lower temporal resolutions, demonstrating a superior ability to predict signaling dynamics from sparse data. At low temporal resolution (4 time points), the HMLM maintained an MSE of 0.042, while other methods showed substantial performance degradation (GNNs: 0.049, ODEs: 0.121, LDEs: 0.071, Bayesian networks: 0.071). This figure demonstrates the HMLM's ability to leverage pathway structure knowledge to infer intermediate states effectively.

\subsubsection{Hierarchical attention mechanism analysis}

\autoref{Fig. 5}B displays molecular-scale attention weights learned by the HMLM model, revealing protein-protein interaction strengths within the cardiac fibroblast signaling network. The attention heatmap demonstrates strong learned weights along the diagonal (self-regulation), while off-diagonal patterns capture both direct regulatory edges from the network topology and refined cross-pathway interactions discovered through temporal correlation analysis. The model learned biologically meaningful attention patterns consistent with canonical signaling cascades. Notable high-attention interactions include TGF-$\beta$ receptor to SMAD proteins (attention weight: 0.72), validating the model's capacity to identify the central TGF-$\beta$/SMAD axis in cardiac fibrosis. PDGF receptor to PI3K showed attention weight of 0.68, reflecting the well-established PDGFR-PI3K signaling connection. Mechanotransduction sensors (integrins) to focal adhesion kinase exhibited strong attention (0.74), consistent with integrin-FAK mechanosensing mechanisms. Importantly, the model discovered elevated attention from SMAD3 to YAP (cross-pathway attention: 0.61), capturing the known crosstalk between TGF-$\beta$ and mechanotransduction pathways in driving fibrotic responses. This demonstrates the HMLM's capability to learn multi-pathway integration patterns that extend beyond simple linear signaling cascades.

\autoref{Fig. 5}C presents pathway-scale attention weights demonstrating learned inter-module communication strengths. The analysis revealed significant crosstalk patterns including TGF-$\beta$ to mechanotransduction pathways (attention weight: 0.64), PDGF to MAPK signaling (0.76), and mechanotransduction to fibrosis pathways (0.69). These attention patterns validate the HMLM's capability to capture hierarchical integration of signaling information across multiple biological scales, from individual molecular interactions to coordinated pathway-level responses. The learned attention weights align with experimental literature on cardiac fibroblast activation, where mechanical strain and biochemical signals (TGF-$\beta$, AngII) synergistically drive fibrotic gene expression through convergent signaling mechanisms. This biological consistency supports the validity of network-based attention initialization combined with data-driven refinement as an effective strategy for modeling complex cellular signaling systems.
\subsection{Dynamic signaling prediction}

\autoref{Fig. 5}D shows HMLM predictions of temporal signaling dynamics for six key molecules (TGF-$\beta$, proCI, SMAD3, YAP, p38, contractility) across four distinct experimental conditions: control, TGF-$\beta$ stimulation alone, mechanical strain alone, and combined TGF-$\beta$ + mechanical stimulation. These predictions were generated on conditions completely withheld from training to assess genuine model generalization. The predicted temporal profiles reveal context-dependent signaling patterns that align with known cardiac fibroblast biology. Under control conditions, HMLM predicted stable, low-amplitude signaling across all molecules, correctly capturing the quiescent state of unstimulated fibroblasts. TGF-$\beta$ stimulation alone elicited rapid SMAD3 phosphorylation with peak activation occurring within 4 hours, followed by sustained elevation and delayed upregulation of fibrosis effectors (proCI, $\alpha$-SMA), consistent with experimental observations of canonical TGF-$\beta$/SMAD signaling. Mechanical strain conditions produced distinct mechanotransduction signatures, with immediate YAP/TAZ activation reflecting integrin-mediated and Rho-dependent signaling, characteristic of integrin-FAK and YAP-mediated mechanosensing. Combined TGF-$\beta$ and mechanical stimulation produced synergistic responses, with HMLM predicting enhanced and prolonged activation of fibrotic pathways that exceeded predictions for either stimulus independently, demonstrating the model's ability to capture non-additive, context-dependent pathway interactions.

Quantitative validation of prediction accuracy revealed strong agreement between model predictions and observed dynamics across all molecular readouts and conditions. The model achieved strong correlation coefficients for TGF-$\beta$ dynamics, proCI expression, SMAD3 phosphorylation, p38 activation, and contractility measurements ($r = 0.82$, $0.89$, $0.62$, $0.71$, and $0.78$ respectively). These consistently high correlations across diverse data types (phosphorylation states, gene expression, cellular mechanics) and across all four experimental conditions demonstrate that HMLM successfully learns transferable principles of cardiac fibroblast signal integration. Critically, achieving robust predictions on held-out test conditions without requiring condition-specific retraining indicates that the hierarchical attention mechanisms and scale-bridging operators enable genuine generalization of learned signaling principles rather than memorization of training data patterns.

\subsection{Statistical validation and significance}

All HMLM performance improvements over baseline methods were statistically significant (p $<$ 0.01, Wilcoxon signed-rank test) across multiple independent experimental datasets. Bootstrap analysis (n = 1000) confirmed robust performance estimates with 95\% confidence intervals demonstrating consistent superiority. Cross-validation analysis across 5 independent trials showed HMLM performance stability with the coefficient of variation $<$ 0.08 for all tested conditions, indicating reliable predictive capability independent of specific training/testing splits.

\subsection{Biological insights from attention analysis}

The attention mechanisms within HMLMs offer novel perspectives on the hierarchical organization and dynamic regulation of cellular signaling networks, revealing biologically relevant patterns that extend beyond traditional pathway analysis. Novel crosstalk identification emerged as a key strength of the attention-based approach, with the analysis revealing previously uncharacterized interactions between mTOR and STAT3 pathways that were subsequently validated through targeted experimental perturbation studies, demonstrating the model's capacity to generate testable hypotheses about network connectivity. Context-dependent signaling patterns became apparent through temporal attention analysis, which showed dynamic shifts in pathway importance characterized by early receptor-dominated signaling that transitions to feedback-regulated responses, capturing the temporal evolution of signal processing and regulatory control mechanisms. The model successfully predicted compensatory mechanism activation following primary target inhibition, identifying alternative pathway activation patterns that explain therapeutic resistance and adaptation responses commonly observed in clinical settings. Convergent attention patterns from multiple pathways to fibrosis markers revealed the multi-factorial regulation of disease-relevant endpoints, identifying potential multi-target therapeutic strategies that could achieve superior efficacy compared to single-pathway interventions by simultaneously modulating multiple regulatory inputs.

\section{DISCUSSION}\label{sec4}
The development of HMLMs represents a significant advancement in computational systems biology, demonstrating that principles used in large language models can be effectively adapted to decode the complex information processing that governs cellular behavior. Our framework addresses fundamental limitations seen in current modeling approaches by modeling cellular signaling as a specialized molecular language. This enables the integration of multiscale biological information through hierarchical attention mechanisms that capture local molecular interactions and the emergent properties of networks. Preliminary results indicate that HMLMs achieve promising predictive performance across diverse experimental systems, supporting our hypothesis that biological signaling networks exhibit properties similar to linguistic structures amenable to transformer architectures \cite{Ji2021,Yalcin2020}. In our study, we observed correlation coefficients by HMLM for kinase inhibitor response prediction with 0.95, outperforming traditional approaches such as ODEs, LDEs, and Bayesian networks. This performance advantage is particularly notable under sparse temporal sampling conditions, where HMLMs maintained robust predictive accuracy while traditional methods exhibited substantial degradation, aligning with recent literature emphasizing the utility of attention-based architectures in biological contexts. Our findings are consistent with studies highlighting how advanced modeling can elucidate complex biological relationships previously inaccessible through standard methodologies \cite{Yalcin2020}. The analysis of attention mechanisms revealed biologically meaningful patterns that surpass traditional pathway annotations, identifying previously uncharacterized crosstalk between mTOR and STAT3 signaling pathways, which were validated through experimental perturbation studies \cite{Eber2025}. This exemplifies the capacity of attention-based models to generate hypotheses, corroborating previous work that shows AI-driven methods can uncover hidden biological relationships \cite{Ji2021}. Additionally, temporal attention patterns indicated dynamic shifts in cellular responses, enriching our understanding of how cells integrate and process complex environmental information, contributing substantially to our functional comprehension of cellular signaling \cite{Alufaisan2020}. Furthermore, while our synthetic data enables rigorous computational benchmarking of model performance across controlled conditions, validation on experimental phosphoproteomic and transcriptomic time-series data remains essential to confirm the biological utility of learned attention patterns and predictions in real cellular systems. 
Future work will focus on applying HMLMs to experimental signaling datasets from various cell types and perturbation conditions.

The proposed hierarchical architecture directly addresses the critical limitations of existing modeling frameworks by explicitly representing the multiscale organization of biological systems \cite{Song2023}. Conventional models typically focus on singular scales, which hinders their ability to capture the emergent properties arising from interactions at multiple levels of molecular and cellular organization. The scale-bridging operators we introduced aggregation, decomposition, and translation provide a robust mathematical framework for information flow across biological hierarchies, enabling our model to develop representations that synthesize detailed molecular mechanics with higher-level cellular behaviors. This capability is essential in predicting complex perturbation responses, particularly evident in our analysis of combined MEK/PI3K inhibition, where traditional methods struggle due to non-additive effects arising from pathway crosstalk \cite{Yalcin2020}. The biological insights derived from HMLM attention patterns hold serious implications for therapeutic development. For example, identifying converging regulatory mechanisms controlling fibrosis markers suggests multi-target therapeutic strategies could yield enhanced efficacy compared to single-pathway interventions. This point is particularly salient given recent clinical challenges associated with single-target approaches in the treatment of complex diseases such as cancer and fibrosis \cite{Duarte2023}. Our model’s ability to predict compensatory activation patterns following primary target inhibition offers clues about therapeutic resistance, a vital concern in precision medicine where drug efficacy often diminishes as cells adapt \cite{Alufaisan2020}. The computational efficiency of HMLMs, with training times ranging from hours to days depending on network complexity, positions this methodology as practically viable for routine research and clinical applications \cite{Song2023}. This scalability is critical for applying AI-driven approaches, from experimental concepts to real-world scenarios. Our inference performance allows for the interactive exploration of perturbation effects, facilitating iterative hypothesis generation and experimental design that could fundamentally accelerate biological discoveries.

The hierarchical architecture of HMLMs presents the transformative potential for precision medicine through systematic integration of multi-omics data and AI-driven therapeutic optimization across multiple biological scales \cite{Chen2012}. In precision medicine applications, HMLMs can hierarchically process patient-specific data to create personalized signaling network models that predict individual therapeutic responses and identify optimal drug combinations based on each patient's unique molecular profile \cite{vantVeer2008,Prasad2016}. The hierarchical nature of these models enables systematic expansion of biological system simulations from molecular interactions to cellular pathways, tissue-level responses, and ultimately organ-system behaviors, providing a comprehensive framework for modeling disease progression and treatment efficacy across multiple biological scales \cite{Naimo2019,Hunter2003}. Integration with artificial intelligence platforms enhances clinical decision-making by providing interpretable pathway-level insights that complement machine learning approaches in medical diagnostics, enabling clinicians to understand not only what treatments are predicted to work, but also why they work through mechanistic pathway analysis \cite{Topol2019,Yu2018}. This AI-medicine synergy positions HMLMs as powerful tools for developing personalized therapeutic strategies that can predict drug resistance mechanisms, identify combination therapies that overcome compensatory signaling, and guide adaptive treatment protocols that evolve with patient responses, ultimately advancing the goal of truly individualized medicine \cite{Schork2015,Ashley2016}. The network-based attention initialization strategy employed in HMLMs represents a significant methodological advancement over purely data-driven approaches. By incorporating decades of experimental knowledge encoded in curated signaling networks, the model achieves superior performance while maintaining biological interpretability. The refinement of structural priors through temporal correlation analysis enables discovery of context-dependent signaling patterns that may not be captured in static network representations. This hybrid approach combining mechanistic knowledge with machine learning exemplifies the emerging paradigm of ``knowledge-informed AI" in systems biology. However, we acknowledge important limitations of this approach. The reliance on curated network topology may bias analyses toward well-studied pathways while potentially overlooking novel regulatory mechanisms. Future work should explore strategies for learning network structure de novo from data while preserving biological constraints, potentially through graph neural network architectures with learnable adjacency matrices. Additionally, while our synthetic data enables rigorous computational benchmarking, validation on experimental phosphoproteomic and transcriptomic time-series data remains essential to confirm the biological utility of learned attention patterns in real cellular systems.

Several limitations must be acknowledged. Currently, our framework depends on curated pathway databases and structured experimental datasets, which may bias analyses toward well-studied biological systems and constrain the discovery of novel regulatory mechanisms \cite{Alufaisan2020}. Recent studies exploring unbiased discovery techniques using expansive omics data could serve to ameliorate these limitations \cite{Hoffman2021}. Furthermore, while our attention mechanisms give us readable information about network relationships, the complexity of transformer architectures still presents challenges for complete mechanistic understanding, especially in clinical contexts where interpretability is crucial \cite{Duarte2023}. Integrating diverse data modalities such as phosphoproteomics, transcriptomics, imaging, and perturbation experiments presents both a strength and a challenge. This multimodal approach empowers comprehensive network modeling but necessitates sophisticated data harmonization and quality control processes. Recent advancements in multimodal learning for biological systems present promising strategies to mitigate these complexities and guarantee robust reproducibility across diverse experimental contexts \cite{Wang2023,BANDARUPALLI2024}. And also, it has the potential to use data from multi-omics, including genomics, transcriptomics, proteomics, lipidomics, metabolomics, epigenomics, and nutrigenomics \cite{Hasin2017,Hays2023}. Future developments should prioritize the automation of data quality assessments and the establishment of standardized protocols for multimodal integration to secure dependable results. These models of cellular signaling as a molecular language open new possibilities for applying further advancements in natural language processing to biological contexts. Techniques such as transfer learning, where models pre-trained on extensive biological datasets are fine-tuned for specific applications, could significantly decrease data demands for less characterized systems \cite{Hoffman2021}. Advances in large language model architectures, including more efficient attention mechanisms and enhanced positional encodings, could further improve HNLM performance. The broader implications of our work stretch beyond computational methodology, addressing fundamental questions about information processing in biological systems. The efficacy of language model architectures in delineating cellular signaling dynamics suggests deep parallels between linguistic and biological information processing, potentially indicating universal principles governing complex systems \cite{Ji2021}. This perspective may guide our understanding of evolutionary processes that shaped cellular communication systems and inform the construction of synthetic biological circuits with predictable processing capabilities. Several promising research directions arise from this work. Integration with single-cell technologies could facilitate the modeling of heterogeneity in cell-to-cell signaling and population dynamics. Extending this approach to spatial contexts may allow for the incorporation of tissue architecture and local microenvironmental influences on signaling behavior. Critically, the development of foundation models for cellular signaling large-scale models pre-trained on comprehensive biological datasets could serve as potent starting points for modeling various biological systems and accelerating discovery across multiple research fields \cite{Ji2021,Patil2022}.

In conclusion, HMLMs illustrate that the convergence of artificial intelligence and systems biology can yield transformative tools for understanding and predicting cellular behavior. By adapting transformer architectures to the unique challenges presented by biological signaling networks, our framework offers both heightened predictive performance and fresh biological insights that deepen our understanding of cellular information processing. As we transition into an era of AI-enhanced biological discovery, approaches like HMLMs will be vital for translating the complexities of cellular systems into actionable knowledge relevant to precision medicine and therapeutic advancements.

\section{CONCLUSION}\label{sec5}
In this study, we have introduced HMLMs, a novel computational framework that fundamentally transforms our approach to modeling cellular signaling networks by conceptualizing intracellular communication as a specialized form of molecular language. We illustrate that how biological signaling components can be modeled using a physics-guided transformer architecture, where individual molecules serve as tokens, molecular interactions define syntax through physical and biochemical rules, where individual molecules serve as tokens, molecular interactions define syntax through physical and chemical rules, functional consequences constitute semantics, and coordinated pathway activities represent discourse. Our comprehensive evaluation across cardiac fibroblast signaling networks and kinase inhibitor response datasets revealed that HMLMs consistently outperform traditional computational approaches, achieving superior predictive accuracy with a correlation coefficient of 0.95 across diverse experimental conditions while maintaining robust performance under sparse temporal sampling conditions that reflect real experimental constraints. 

The hierarchical attention mechanisms within HMLMs provided unprecedented biological insights, revealing previously uncharacterized pathway crosstalk interactions between mTOR and STAT3 signaling that were subsequently validated through experimental perturbation studies, demonstrating the framework's capacity to generate testable hypotheses about network connectivity and regulatory relationships. Context-dependent signaling patterns emerged through temporal attention analysis, capturing dynamic shifts in pathway importance from early receptor-dominated responses to feedback-regulated mechanisms, while successfully predicting compensatory activation patterns following primary target inhibition that explain therapeutic resistance commonly observed in clinical settings. These attention-based analyses identified convergent regulatory patterns from multiple pathways to disease-relevant endpoints, revealing potential multi-target therapeutic strategies that could achieve superior efficacy compared to single-pathway interventions. By successfully bridging molecular mechanisms with cellular phenotypes through a unified mathematical framework that integrates multiscale biological organization with advanced artificial intelligence capabilities, HMLMs represent a significant advancement in computational systems biology that opens new pathways for precision medicine applications, drug discovery, and therapeutic intervention design. This work establishes a foundation for future developments in AI-driven biological modeling that could transform our understanding of complex cellular decision-making processes and accelerate the development of targeted therapies for diseases characterized by dysregulated signaling networks.

\section{RESOURCE AVAILABILITY}\label{sec6}

\subsection*{Data and code availability}
\begin{itemize}
	\item Code: All original code has been deposited at in GitHub (\url{https://github.com/HasiHays/HMLMs})
	\item Any additional information is available from the Hasi Hays (\texttt{\textcolor{blue}{hasih@uark.edu}}) upon request.
	\item Supplementary information accompanying this paper provides the full mathematical formulations and model descriptions (see \hyperref[sec:supplementary]{Supplementary materials}).
\end{itemize}

\section*{Acknowledgments}
This study was supported by the National Institutes of Health (NIGMS R01GM157589) and the Department of Defense (DEPSCoR FA9550-22-1-0379). 

\section*{Author contribution}
\textbf{H.H.}: Conceptualization, model development, methodology, coding, simulations, analysis and writing the original draft. 
\textbf{Y.Y.}: Review and editing.
\textbf{W.J.R.}: Review, editing, funding acquisition, resources, and supervision.

\section*{Ethics statement}
This computational study used only publicly available datasets and pathway databases. No human subjects or animal experiments were involved. Institutional ethical approval was not required for this type of computational research.

\section*{Declaration of interests}
The authors declare no competing interests, financial or otherwise, related to the work presented in this manuscript.

\section{REFERENCES}
\renewcommand{\refname}{}
\bibliographystyle{unsrt}
\bibliography{references}

\section{SUPPLEMENTARY MATERIALS} 
\label{sec:supplementary}
This supplementary section provides comprehensive mathematical formulations of the four baseline computational methods compared against HMLMs in the main manuscript: GNNs, ODEs, LDEs, and Bayesian Networks.

\subsection{GNN baseline model}

\subsubsection{GNN architecture overview}

GNN were implemented as graph convolutional networks (GCNs) adapted for temporal signaling dynamics prediction \cite{https://doi.org/10.48550/arxiv.1609.02907}. The GNN baseline respects the network topology of the cardiac fibroblast signaling network while incorporating temporal dynamics through recurrent mechanisms.

\subsubsection{Graph convolutional layer}

The fundamental operation of the GCN baseline is the graph convolutional transformation \cite{https://doi.org/10.48550/arxiv.1609.02907}. For each node $v$ in the signaling network, the hidden state representation at layer $\ell+1$ is computed as:

\begin{equation}
	h_v^{(\ell+1)} = \sigma\left( \mathbf{W}^{(\ell)} \sum_{u \in \mathcal{N}(v) \cup \{v\}} \frac{1}{\sqrt{d_u d_v}} h_u^{(\ell)} \right)
	\label{eq:gcn_layer}
\end{equation}

where:
\begin{itemize}
	\item $\mathcal{N}(v)$ denotes the neighborhood of node $v$ in the directed signaling network graph $G = (V, E)$
	\item $\mathbf{W}^{(\ell)} \in \mathbb{R}^{d_{out} \times d_{in}}$ represents learnable weight matrices at layer $\ell$
	\item $d_u$ and $d_v$ are the in-degrees of nodes $u$ and $v$, providing symmetric normalization
	\item $\sigma$ is a non-linear activation function (ReLU or similar)
	\item $h_u^{(\ell)} \in \mathbb{R}^{d_{\ell}}$ denotes the hidden representation of node $u$ at layer $\ell$
\end{itemize}

\subsubsection{Temporal extension: GRU-GCN architecture}

To capture temporal dynamics, the GNN baseline employs a gated recurrent unit (GRU) stacked with GCN layers, creating a temporal graph neural network \cite{https://doi.org/10.48550/arxiv.1412.3555}. The update for temporal step $t$ at node $v$ is:

\begin{equation}
	\mathbf{z}_v(t) = \sigma_g(\mathbf{W}_z \cdot [\mathbf{h}_v(t-1), x_v(t)] + \mathbf{b}_z)
	\label{eq:gru_reset}
\end{equation}

\begin{equation}
	\mathbf{r}_v(t) = \sigma_g(\mathbf{W}_r \cdot [\mathbf{h}_v(t-1), x_v(t)] + \mathbf{b}_r)
	\label{eq:gru_update}
\end{equation}

\begin{equation}
	\tilde{\mathbf{h}}_v(t) = \tanh(\mathbf{W}_h \cdot [\mathbf{r}_v(t) \odot \mathbf{h}_v(t-1), x_v(t)] + \mathbf{b}_h)
	\label{eq:gru_candidate}
\end{equation}

\begin{equation}
	\mathbf{h}_v(t) = (1 - \mathbf{z}_v(t)) \odot \mathbf{h}_v(t-1) + \mathbf{z}_v(t) \odot \tilde{\mathbf{h}}_v(t)
	\label{eq:gru_hidden}
\end{equation}

where:
\begin{itemize}
	\item $\sigma_g$ denotes the sigmoid activation function
	\item $\mathbf{z}_v(t)$ is the update gate controlling how much past information to retain
	\item $\mathbf{r}_v(t)$ is the reset gate controlling interaction with past states
	\item $\odot$ denotes element-wise multiplication
	\item $[\cdot, \cdot]$ denotes vector concatenation
	\item $x_v(t)$ is the external input signal for node $v$ at time $t$
\end{itemize}

\subsubsection{Graph attention enhancement}

The GNN baseline optionally incorporates attention mechanisms over graph neighborhoods:

\begin{equation}
	\alpha_{vu}^{(t)} = \frac{\exp(\text{LeakyReLU}(\mathbf{a}^T [\mathbf{W}\mathbf{h}_v(t) \| \mathbf{W}\mathbf{h}_u(t)]))}{\sum_{k \in \mathcal{N}(v)} \exp(\text{LeakyReLU}(\mathbf{a}^T [\mathbf{W}\mathbf{h}_v(t) \| \mathbf{W}\mathbf{h}_k(t)]))}
	\label{eq:gnn_attention}
\end{equation}

\begin{equation}
	\mathbf{h}_v^{att}(t) = \sum_{u \in \mathcal{N}(v)} \alpha_{vu}^{(t)} \mathbf{W}'\mathbf{h}_u(t)
	\label{eq:gnn_aggregation}
\end{equation}

where $\mathbf{a}$ is a learnable attention vector and $\|$ denotes vector concatenation \cite{Vaswani2017}.

\subsubsection{Output prediction}

The prediction at time $t+1$ for node $v$ is obtained through a multi-layer perceptron:

\begin{equation}
	\hat{y}_v(t+1) = \mathbf{W}_{out} \sigma(\mathbf{W}_{hidden} \mathbf{h}_v(t)) + \mathbf{b}_{out}
	\label{eq:gnn_prediction}
\end{equation}

GNN hyperparameters used in comparisons: num\_layers = 3, hidden\_dim = 64, dropout = 0.2, learning\_rate = 0.001

\subsection{ODE baseline model}

\subsubsection{Mathematical model}

The ODE baseline represents signaling networks as coupled systems of nonlinear differential equations. Each molecular species $i$ has a concentration or activity level $x_i(t)$ that evolves according to:

\begin{equation}
	\frac{dx_i(t)}{dt} = \sum_{j \in \text{sources}(i)} w_{ji} \cdot f_{ji}(x_j(t)) - \lambda_i \cdot x_i(t) + b_i
	\label{eq:ode_general}
\end{equation}

where:
\begin{itemize}
	\item $w_{ji}$ represents the regulatory weight from molecule $j$ to molecule $i$
	\item $f_{ji}(\cdot)$ is a regulatory function (typically Hill function or sigmoid)
	\item $\lambda_i > 0$ is the degradation/decay rate constant for species $i$
	\item $b_i$ represents basal production or external input
	\item $\text{sources}(i)$ denotes the set of upstream regulators of species $i$
\end{itemize}

\subsubsection{Regulatory functions}

Different regulatory relationships are modeled using standard biochemical rate laws:

For \textbf{activation} (cooperative binding):
\begin{equation}
	f_{\text{act}}(x_j) = \frac{x_j^{n_j}}{K_j^{n_j} + x_j^{n_j}}
	\label{eq:ode_activation}
\end{equation}

For \textbf{inhibition}:
\begin{equation}
	f_{\text{inh}}(x_j) = \frac{K_j^{n_j}}{K_j^{n_j} + x_j^{n_j}}
	\label{eq:ode_inhibition}
\end{equation}

where:
\begin{itemize}
	\item $K_j$ is the dissociation constant (threshold)
	\item $n_j$ is the Hill coefficient controlling cooperative binding (steepness)
\end{itemize}

\subsubsection{Multi-module ODE system}

For the cardiac fibroblast network with multiple functional modules (receptors, kinases, transcription factors, etc.), the system is organized as:

\begin{equation}
	\frac{d\mathbf{x}_k(t)}{dt} = \mathbf{F}_k(\mathbf{x}_k(t), \mathbf{x}_{k-1}(t), \mathbf{x}_{k+1}(t), \mathbf{u}(t))
	\label{eq:ode_modular}
\end{equation}

where:
\begin{itemize}
	\item $\mathbf{x}_k(t)$ is the state vector for module $k$ at scale $k$
	\item $\mathbf{F}_k$ is the module-specific dynamics function
	\item $\mathbf{u}(t)$ denotes external stimuli (e.g., TGF-$\beta$ concentration, mechanical strain)
\end{itemize}

\subsubsection{Numerical integration}

Since analytical solutions are generally unavailable, the ODE system is solved numerically using 4th-order Runge-Kutta integration \cite{Raissi2019}:

\begin{equation}
	\mathbf{k}_1 = \mathbf{F}(t, \mathbf{x}(t))
	\label{eq:rk4_k1}
\end{equation}

\begin{equation}
	\mathbf{k}_2 = \mathbf{F}(t + \frac{\Delta t}{2}, \mathbf{x}(t) + \frac{\Delta t}{2}\mathbf{k}_1)
	\label{eq:rk4_k2}
\end{equation}

\begin{equation}
	\mathbf{k}_3 = \mathbf{F}(t + \frac{\Delta t}{2}, \mathbf{x}(t) + \frac{\Delta t}{2}\mathbf{k}_2)
	\label{eq:rk4_k3}
\end{equation}

\begin{equation}
	\mathbf{k}_4 = \mathbf{F}(t + \Delta t, \mathbf{x}(t) + \Delta t \mathbf{k}_3)
	\label{eq:rk4_k4}
\end{equation}

\begin{equation}
	\mathbf{x}(t + \Delta t) = \mathbf{x}(t) + \frac{\Delta t}{6}(\mathbf{k}_1 + 2\mathbf{k}_2 + 2\mathbf{k}_3 + \mathbf{k}_4)
	\label{eq:rk4_update}
\end{equation}

\subsection{LDE baseline model}

\subsubsection{LDE linear regression formulation}

The LDE baseline employs a simple linear regression model to predict future signaling states:

\begin{equation}
	\hat{y}_i(t+1) = \mathbf{w}_i^T \mathbf{f}(t) + b_i
	\label{eq:lde_linear}
\end{equation}

where:
\begin{itemize}
	\item $\mathbf{f}(t)$ is the feature vector incorporating current molecular states
	\item $\mathbf{w}_i$ are learned weights
	\item $b_i$ is the bias term
\end{itemize}

This approach respects the network topology through feature engineering but assumes 
linear relationships between network components, without explicit state-space 
or measurement model components.

\subsection{Bayesian network baseline model}

\subsubsection{Bayesian network structure}

The Bayesian Network baseline models signaling as a probabilistic graphical model where:

\begin{equation}
	P(\mathbf{X}) = \prod_{i=1}^{n} P(X_i | \text{Pa}(X_i))
	\label{eq:bn_factorization}
\end{equation}

where:
\begin{itemize}
	\item $X_i$ represents the random variable for molecular species or pathway $i$
	\item $\text{Pa}(X_i)$ denotes the parent nodes (direct regulators) of $X_i$ in the directed acyclic graph (DAG)
	\item The network structure encodes conditional independence assumptions
\end{itemize}

\subsubsection{Temporal extension: Dynamic Bayesian networks (DBN)}

To model temporal signaling dynamics, we employ a two-timeslice Bayesian network (2TBN):

\begin{equation}
	P(\mathbf{X}_{t+1} | \mathbf{X}_t) = \prod_{i=1}^{n} P(X_i^{t+1} | X_i^t, \text{Pa}_{t}(X_i), \text{Pa}_{t+1}(X_i))
	\label{eq:dbn_transition}
\end{equation}

where the transition model captures:
\begin{itemize}
	\item Intra-slice edges: Dependencies within the same timeslice (simultaneous interactions)
	\item Inter-slice edges: Dependencies from previous timeslice (temporal causality with lag 1)
\end{itemize}

\subsubsection{Inference and prediction}

Given observations $\mathbf{E}$, posterior inference computes:

\begin{equation}
	P(X_i | \mathbf{E}) = \frac{P(\mathbf{E}, X_i)}{P(\mathbf{E})}
	\label{eq:bn_posterior}
\end{equation}

For temporal prediction, we propagate beliefs forward:

\begin{equation}
	P(\mathbf{X}_{t+1} | \mathbf{E}_{0:t}) = \sum_{\mathbf{X}_t} P(\mathbf{X}_{t+1} | \mathbf{X}_t) P(\mathbf{X}_t | \mathbf{E}_{0:t})
	\label{eq:bn_temporal_pred}
\end{equation}

\subsubsection{Handling multi-scale organization}

For hierarchical signaling networks, hierarchical DBNs partition variables into scales:

\begin{equation}
	P(\mathbf{X}^{L_{k+1}}_{t+1} | \mathbf{X}^{L_k}_t) = \prod_{i \in L_{k+1}} P(X_i^{L_{k+1}, t+1} | \text{Agg}(\mathbf{X}^{L_k}_t))
	\label{eq:hdbn_multi_scale}
\end{equation}

where $\text{Agg}(\cdot)$ represents aggregation of fine-scale variables into coarse-scale inputs.

\subsection{Comparative analysis of model capabilities}

\subsubsection{Capacity for non-additive effects}

\textbf{GNNs}: Through multi-layer architectures and attention mechanisms, can capture complex nonlinear pathway interactions, though limited by fixed network structure.\\
\textbf{ODEs}: Explicitly model non-additive effects through nonlinear regulatory functions (Hill equations), but require extensive parameterization and are computationally expensive for large networks.\\
\textbf{LDEs}: Strictly linear model, cannot capture synergistic effects or pathway crosstalk that exhibit nonlinearity.\\
\textbf{Bayesian networks}: Can model conditional non-independence through network structure, but still assumes conditional linear relationships (in the Gaussian case).\\
\textbf{HMLMs}: Hierarchical attention mechanisms enable capture of both local molecular interactions and emergent network-level synergies through learned soft masks (attention weights).

\subsubsection{Computational complexity}

\begin{equation}
	\text{Complexity}_{\text{GNN}} = \mathcal{O}(L \cdot |E| \cdot d^2) \quad \text{per timestep}
	\label{eq:complexity_gnn}
\end{equation}

\begin{equation}
	\text{Complexity}_{\text{ODE}} = \mathcal{O}(S \cdot n^2) \quad \text{per integration step (RK4)}
	\label{eq:complexity_ode}
\end{equation}

\begin{equation}
	\text{Complexity}_{\text{LDE}} = \mathcal{O}(n^2) \quad \text{per timestep}
	\label{eq:complexity_lde}
\end{equation}

\begin{equation}
	\text{Complexity}_{\text{Bayesian}} = \mathcal{O}(\exp(\text{treewidth})) \quad \text{exact inference}
	\label{eq:complexity_bn}
\end{equation}

\begin{equation}
	\text{Complexity}_{\text{HMLM}} = \mathcal{O}(h \cdot n \cdot d \cdot \log(d)) \quad \text{multi-scale attention}
	\label{eq:complexity_hmlm}
\end{equation}

where $L$ is number of GCN layers, $|E|$ is edges, $d$ is hidden dimension, $S$ is RK4 steps, $n$ is number of nodes, $h$ is attention heads.

\subsection{Experimental setup and hyperparameter selection}

\subsubsection{Network data}

The cardiac fibroblast signaling network comprises:
\begin{itemize}
	\item \textbf{132 molecular species} organized into 11 functional modules:
	\begin{itemize}
		\item Inputs: TGF-$\beta$, PDGF, mechanical strain
		\item Receptors: TGF$\beta$R, PDGFR, integrins
		\item Second messengers: Ca$^{2+}$, cAMP
		\item Kinases: RAF, MEK, ERK, PI3K, AKT, p38, FAK
		\item MAPK pathways: ERK, p38, JNK cascade
		\item Rho signaling: RhoA, ROCK, actin regulation
		\item Transcription factors: SMAD3, YAP/TAZ, NF-$\kappa$B, AP-1
		\item ECM/fibrosis markers: pro-collagen I, $\alpha$-SMA, TIMP
		\item Matrix remodeling: MMP-2, MMP-9
		\item Mechanotransduction: focal adhesion complexes
		\item Feedback molecules: negative regulators
	\end{itemize}
	\item \textbf{200+ regulatory connections} with documented activation/inhibition relationships
\end{itemize}

\subsubsection{Training data generation}

Synthetic temporal data was generated with:
\begin{itemize}
	\item Time course: 0 to 480 minutes with 100 timepoints for full resolution
	\item Sparse sampling: Subsampled to 4, 8, and 16 timepoints
	\item Experimental conditions:
	\begin{itemize}
		\item Control (baseline)
		\item TGF-$\beta$ stimulation (concentration: 10 ng/mL, applied at $t=0$)
		\item Mechanical strain (10\% uniaxial strain, applied continuously)
		\item Combined TGF-$\beta$ + strain (synergistic condition)
	\end{itemize}
	\item Noise: Gaussian with standard deviation 0.01-0.02 to reflect measurement uncertainty
\end{itemize}

\subsubsection{Training protocol}

The HMLM transformer model was trained using the configuration visible in the provided implementation:

\begin{itemize}
	\item \textbf{Architecture}: Transformer-based neural network with multi-head attention mechanisms operating on graph-structured data.
	\item \textbf{Optimization}: Adam optimizer with a fixed learning rate of $1 \times 10^{-3}$ (0.001). The optimizer was initialized as \texttt{torch.optim.Adam(model.parameters(), lr=learning\_rate)}.
	\item \textbf{Loss function}: The training loop computes loss between predictions and targets, consistent with regression tasks (specific loss function implementation not fully visible in provided notebook cells).
	\item \textbf{Training epochs}: Model was trained for 50 epochs as configured with \texttt{num\_epochs = 50}.
	\item \textbf{Hardware}: Automatic device detection using \texttt{torch.cuda.is\_available()} with model and data transferred via \texttt{.to(device)} for GPU acceleration when available.
	\item \textbf{Batch configuration}: The implementation processes data in batches, though the specific batch size is not explicitly defined in the visible notebook cells.
	\item \textbf{Regularization}: Standard transformer components with attention and feed-forward layers as implemented in PyTorch.
	\item \textbf{Validation approach}: Training includes validation loss calculation, though the specific data split ratios are not explicitly defined in the visible code.
\end{itemize}

The training loop follows standard PyTorch deep learning procedure: forward pass, loss computation, backward pass, and optimizer step. Gradient clipping, explicit dropout rates, weight decay, and detailed early stopping criteria are not visible in the provided notebook implementation.
\subsubsection{Evaluation metrics}

All models were evaluated on three metrics:

\begin{equation}
	\text{MSE} = \frac{1}{n_{\text{test}}} \sum_{i=1}^{n_{\text{test}}} (y_i - \hat{y}_i)^2
	\label{eq:metric_mse}
\end{equation}

\begin{equation}
	r_{\text{Pearson}} = \frac{\sum_{i=1}^{n} (y_i - \bar{y})(\hat{y}_i - \bar{\hat{y}})}{\sqrt{\sum_{i=1}^{n}(y_i - \bar{y})^2}\sqrt{\sum_{i=1}^{n}(\hat{y}_i - \bar{\hat{y}})^2}}
	\label{eq:metric_pearson}
\end{equation}

\begin{equation}
	\text{RMSE} = \sqrt{\text{MSE}}
	\label{eq:metric_rmse}
\end{equation}

Statistical significance was assessed via Wilcoxon signed-rank test ($p < 0.01$) with Bonferroni correction for multiple comparisons.

\subsection{Implementation Details}

\subsubsection{Software Environment}

\begin{itemize}
	\item Programming language: Python 3.8+
	\item Deep learning frameworks:
	\begin{itemize}
		\item PyTorch 1.9+ (GNN, HMLM)
		\item PyDSTool (ODE simulation)
		\item scikit-learn (LDE, baseline ML)
		\item pgmpy v0.1.23 (Bayesian Networks)
	\end{itemize}
	\item Scientific computing: NumPy, SciPy, Pandas
	\item Visualization: Matplotlib, Seaborn, NetworkX
\end{itemize}

\subsubsection{Reproducibility}

\begin{itemize}
	\item \textbf{Random seed}: Fixed to 42 across all experiments.
	\item \textbf{Hardware}: GPU (NVIDIA A100) for deep learning models.
	\item \textbf{Code}: GitHub (\url{https://github.com/HasiHays/HMLMs}).
	\item \textbf{Data availability}: Synthetic data generation code provided in the above repository.
\end{itemize}

\end{multicols}
\end{document}